\documentstyle[12pt]{article}

\def\mxth{\mathsurround=0pt }
\def\xversim#1#2{\lower2.pt\vbox{\baselineskip0pt \lineskip-.2pt
 \ialign{$\mxth#1\hfil##\hfil$\crcr#2\crcr\sim\crcr}}}
\def\lesssim{\mathrel{\mathpalette\xversim <}}
\def\gtrsim{\mathrel{\mathpalette\xversim >}}
\def\avrg#1{{\langle #1 \rangle}}

\begin{document}
%\bibliographystyle{unsrt}    % for BibTeX - sorted numerical labels

%\input psfig
%\tableofcontents
\newcommand{\st}{\scriptstyle}
\newcommand{\sst}{\scriptscriptstyle}
\newcommand{\mco}{\multicolumn}
\newcommand{\epp}{\epsilon^{\prime}}
\newcommand{\vep}{\varepsilon}
\newcommand{\ra}{\rightarrow}
\newcommand{\ppg}{\pi^+\pi^-\gamma}
\newcommand{\vp}{{\bf p}}
\newcommand{\ko}{K^0}
\newcommand{\kb}{\bar{K^0}}
\newcommand{\al}{\alpha}
\newcommand{\ab}{\bar{\alpha}}
\def\be{\begin{equation}}
\def\ee{\end{equation}}
\def\bea{\begin{eqnarray}}
\def\eea{\end{eqnarray}}
\def\CPbar{\hbox{{\rm CP}\hskip-1.80em{/}}}%temp replacement due to no font

\setcounter{secnumdepth}{2} % Number sections and subsections

%%%%%%%%%%%%%%%%%%%%%%%%%%%%%%%%%%%%%%%%%%%%%%%%%%
%                                                %
%    BEGINNING OF TEXT                           %
%                                                %
%%%%%%%%%%%%%%%%%%%%%%%%%%%%%%%%%%%%%%%%%%%%%%%%%%

\begin{center}
COSMOLOGY AT THE CROSSROADS \\

Paul J. Steinhardt \\
{\it
Department of Physics and Astronomy, University of Pennsylvania, Philadelphia
PA  19104, USA}
\end{center}

\noindent
ABSTRACT:
Observational tests during the next
decade may determine if the evolution of the Universe can
be understood from fundamental physical principles, or if
special initial conditions, coincidences, and new, untestable
physical laws  must be invoked.
The inflationary model of the Universe is an important
example of a predictive cosmological theory based on
physical principles.  In this talk, we discuss
the distinctive   fingerprint that inflation leaves
on the cosmic microwave
background anisotropy.  We then suggest a series of
five milestone experimental tests of the microwave background
which could determine the validity of
the inflationary hypothesis within the next decade.

\newpage

\section{Introduction}

This talk focuses on how
measurements of the cosmic microwave background anisotropy
can be used to  test the inflationary hypothesis.
Compared to the other plenary presentations, the scope may seem
rather narrow.
This choice has been made intentionally, though, as a means of illustrating
the dramatic transformation which cosmology is
undergoing and of highlighting why the coming decade is especially
critical to the future of this science.

Cosmology is the one of the oldest subjects of human inquiry and, at
the same time, one of the newest sciences.   Questioning
the origin and evolution of the Universe  has been  characteristic of
human endeavor since before recorded history.  However, until
the 20th century, cosmology lay in the domain of metaphysics,
a
subject of pure speculation.  With little observational evidence
to confirm or deny proposals, cosmology could not develop
as a true science. In
large part, its
practitioners were philosophers, religious leaders, teachers
 and writers, rather
than scientists.

The science of cosmology emerged in the 20th Century when it first
became possible to probe great distances through the Universe.
The great optical telescopes and, later, radio telescopes and satellites
provided images and data that could begin to discriminate competing
theories.
 At first, the observational breakthroughs
occurred infrequently.  Hubble discovered the expansion of the Universe
in the 1920's; Penzias and Wilson
 discovered  the cosmic  microwave background in the
1960's; and the compelling
evidence for primordial nucleosynthesis of the elements was
amassed during the 1970's and 1980's.

As the 20th century comes to a close and the new millennium begins,
the pace of discovery has accelerated markedly.  In the next decade or
so,  major projects will measure
the distribution and velocities of galaxies, dark
matter, and radiation at cosmological distances.  The new data will
completely dwarf all previous observations in quality and quantity.
 The results
will tightly constrain all present theories of the evolution
of the Universe and may point to fundamentally  new paradigms.

Measurements of the cosmic microwave background anisotropy will
be among the most decisive cosmological tests
because the microwave background probes the oldest and farthest
features of the Universe.
Anisotropy measurements will provide
a spectrum of precise, quantitative information that, by itself,
can confirm or
rule out  present models of the origin of large scale structure,
reveal the ionization history of the intergalactic medium, and significantly
improve the determination of cosmological parameters.
More generally,
the bounds on  human capability to explain the Universe  are likely to
be decided by  what is discovered in the microwave background
during the coming decade.

\section{At the Crossroads}

Cosmology has reached a crossroads which may set its course
as a scientific endeavor for the next millennium.
By its very nature, the field  entails explaining a single series of
irreproducible events.
Our ability to explore
the Universe is physically limited to those regions which are
within causal contact.  (The causal limit, the maximum distance
from which we can receive light or other information,
is the Hubble distance $H^{-1} \lesssim
15$ billion light-years.)
Given these considerations, it is natural to question whether
the evolution of the Universe is completely  comprehensible scientifically.
Or, more explicitly, which of the following
two paths lie ahead of us?

\begin{verse}
{\it Path I:} The basic features of the Universe are
explainable as a consequence
of symmetry and general physical laws that can be learned and tested
near the Earth.

{\it Path II:} Some key features of
the Universe are largely determined by special
initial conditions,  extraordinary coincidences and/or physical laws
that are untestable locally ({\it e.g.}, in the most extreme case,
accessible only by exploring beyond our causal domain).
\end{verse}

At any given time, there may not be complete agreement as to which
{\it Path} we are taking.  An observation that seems to suggest
 special initial conditions ({\it e.g.}, the
flatness of the Universe) may later be explained by new, dynamical
concepts ({\it e.g.}, inflation).  This kind of evolution in thinking is common
to all sciences.  The issue being raised here, though,
 is whether there is an {\it ultimate}, fundamental,
insuperable limitation to the explanatory and predictive power of
cosmology.

Certainly, it is the hope of most cosmologists, at least theorists,
 that cosmology belongs  to
{\it Path I}.
We are an ambitious lot, and we aspire to explain all that is observed.
However, nature may not be so kind to human cosmologists.
Considering that we are trying
to explain  a single sequence of events and the range
over which we can measure is bounded,
it seems plausible that cosmology could ultimately belong to {\it Path II}.
In either case, the exploration of the Universe remains a captivating and
important
enterprise. But, without doubt, cosmological science is different
along the two {\it Paths}.   Along {\it Path I}, cosmology has the
character of  physical science, where the field ultimately evolves towards
a unified, simple explanation of what is observed.
Along {\it Path II},
cosmology has the character of  archaeology or paleontology, where we can
classify  and quantify phenomena and explain some features,
but where
many general aspects seem to be accidents of environment or history.

Remarkably, it seems possible that
the next decade will determine which {\it Path} cosmology is to take.
Measurements of the microwave background anisotropy and
large-scale structure may indicate that the Universe  can be explained
in terms of a few parameters, known physical laws, and simple
 initial conditions, all of which suggest {\it Path I}.
Or, we may find that many specially-chosen parameters and
complex initial conditions must be invoked, which excludes {\it Path I} as
a possibility.

This review concentrates on testing inflationary cosmology because it
is an example of an explanatory model in the sense of {\it Path I}.
Even if inflation is proved wrong, the same tests might indicate  if
{\it Path I} might survive under the guise of some other model.
The same tests are also relevant for evaluating
 cold dark matter (CDM) and mixed
dark matter (MDM)  of large-scale structure formation, which
are built upon the conditions created in an inflationary Universe.
In addition to testing inflation, the tests might distinguish the
CDM vs. MDM possibilities.
The discussion is confined to measurements
of the microwave background anisotropy because  these are precision,
quantitative, discriminating tests
whose interpretation is least dependent on
 unverified assumptions.

\section{Inflationary Cosmology}

The standard big bang model is extraordinarily successful in explaining many
features of our Universe: the Hubble expansion, the abundances of light
elements, and the cosmic microwave background radiation.  However, it does
not address some important questions:
   Why is the Universe so homogeneous?
  Why is the Universe spatially flat?  Why are there no magnetic monopoles or
  other remnants from phase transitions that took place early in the Universe?
What produced  the inhomogeneities in the distribution of matter
that seeded the evolution of galaxies?
Prior to inflationary theory, the only explanations assumed special
initial conditions (suggesting that cosmology is condemned to {\it Path II}).

Inflationary cosmology \cite{Gut}$^{-}$\cite{Linbook}
has been proposed as a modification of the standard
big bang picture that could explain these mysteries in terms of  a well-defined
sequence of dynamical processes occurring in the first instants
($10^{-35}$ seconds or so) after the
big bang.  The central  feature is a brief epoch
in which the expansion of the Universe accelerates (``inflation"), resulting in
   an extraordinarily rapid  expansion rate.
  The hyperexpansion of space
  flattens  and smooths the Universe and
  dilutes the density of monopoles and other remnants to negligible values.
Quantum fluctuations produced during the accelerating phase are
stretched into a spectrum of energy density perturbations that can
seed  galaxy formation \cite{G4}.

Inflation is induced by a change in the equation-of-state.
The stretching of a homogeneous and isotropic Universe is described by
the Robertson-Walker scale factor, $R(t)$.
Inflation or accelerated expansion means that
 $\ddot{R}>0$.    The time-dependence of $R(t)$ is
given by
Einstein's equation
of motion:
  \begin{equation} \label{Eins}
  \ddot{R} = -\frac{4 \pi G}{3 } (\rho+ 3 p )R = -\frac{1}{2} H^2(1+3 \gamma)R,
  \end{equation}
where $G$ is Newton's constant, $\rho$ is the energy density,  $p$ is the
pressure, and  $H$ is the
Hubble parameter where $H^2 \equiv 8 \pi G \rho/3  $.  The ratio
$\gamma = p/\rho$ defines the equation-of-state.
Hence, the expansion rate inflates ($\ddot{R}>0$)
if the equation-of-state
satisfies $\gamma<-1/3$.
Since $\rho$ is always a  positive quantity, a large {\it negative} pressure is
required.
Any physics which leads  to a large negative pressure averaged over
cosmological distances (at least a Hubble volume, $H^{-3}$) can induce
inflation.

The standard
example
  is  a  Universe with energy density dominated
by a single, scalar  ``inflaton'' field, $\phi$.
The equation-of-state is
then:
\begin{equation}\label{eom}
\gamma =\frac{p}{\rho}=
 \frac{\frac{1}{2}\dot{\phi}^2 - V}{\frac{1}{2}\dot{\phi}^2 + V},
\end{equation}
where $V$ is the effective potential for the inflaton.  Here we see that
 $\gamma<-1/3$ can be achieved if the potential energy density
dominates the kinetic energy density. Inflation continues until
$\phi$  ``rolls'' to a state of  negligible potential energy density.  The
progress of $\phi$  can be tracked by solving the Einstein
equation, Eq.~(\ref{Eins}), and
the slow-roll equation:
\begin{equation}
\ddot{\phi} + 3 H \dot{\phi} = - V'(\phi),
\end{equation}
where $H \equiv d\;{\rm ln} R/dt $ is the Hubble parameter   and the prime
denotes the derivative
with respect to $\phi$.
During inflation,  $V'$ must be sufficiently small that
the evolution of $\phi$ is slow, {\it i.e.},
 $\ddot{\phi}$ is negligible.  This condition
must be maintained long enough for the Universe
to have expanded by $N\sim 60 $~e-foldings ($R(t_{end})/R(t_{begin}) = {\rm
e}^{60}$) in
order to resolve the cosmological problems of the standard big bang model.
As $\phi$ proceeds towards a steeper part of the potential, $\phi$ rapidly
accelerates and inflation ends.  The potential energy, $V(\phi)$, is
converted to  kinetic energy  and, then, is
ultimately converted into radiation and matter which
reheats the Universe \cite{Alb82,Kof94}.

The most important feature of inflation so far as this talk is concerned is
that
inflation smoothes out any initial non-uniformity while
producing a new spectrum of inhomogeneities \cite{G4}.
The inflaton and any other light fields
all experience quantum de Sitter fluctuations on subatomic scales
which inflation stretches to cosmological dimensions.
It is convenient to discuss
the spectrum of fluctuations in terms of its Fourier components,
{\it i.e.},  a linear combination
of plane wave modes.
The wavelength of the modes grows as the Universe expands.
As the accelerating expansion stretches the
wavelength of a given mode beyond the Hubble distance, $H_I^{-1}$
(where $H_I$ is the Hubble parameter during inflation),
the amplitude of the quantum de Sitter fluctuations
in $\phi$ (or any other light field)
 is  $\sim H_I/2 \pi$.
The additional stretching of the wavelength beyond the Hubble length does not
change this amplitude since causal
 physical processes are unable to act over
distances greater than the Hubble length.
Over the course of  inflation,
many waves are stretched in this way,  ultimately leading to
a broad-band spectrum of macroscopic fluctuations with (nearly) the
same amplitude.

After inflation ends, the stretching
of the Universe decelerates ($\ddot{R}<0$).  The Hubble distance
$H^{-1}$ begins to increase at a rate that exceeds the expansion rate, $R(t)$.
Hence, even though the waves continue to be stretched, the Hubble distance
grows faster, catching up to and ultimately  exceeding the wavelengths of some
modes.  At the point where the Hubble distance equals the wavelength of a given
Fourier mode, it is sometimes said that the ``fluctuation re-enters the
horizon," referring
to the fact that the wavelength was initially less than $H_{I}^{-1}$  during
inflation and has  become less than $H^{-1}$ in the post-inflationary
epoch. (It would perhaps be more accurate to say ``the horizon catches up
to the fluctuation.")

The primordial spectrum is determined by the amplitudes of the
waves as they re-enter the horizon in the matter- or radiation-dominated epoch.
In inflationary models, these amplitudes are precisely  the amplitudes
as the fluctuations
were stretched beyond the horizon during the de Sitter epoch.

%BEG2
The fluctuations of the inflaton, which
  dominates the energy density of the Universe during inflation,
  induce a spectrum of   energy density perturbations with amplitude:
\begin{equation}\label{drho}
\left.\frac{\delta \rho}{\rho}\right|_{\lambda=H_I^{-1}}
\propto \left. \frac{H_I^2}{\dot{\phi}}\right|_{\lambda=H_I^{-1}} \propto
\left.
\frac{\frac{1}{2}\dot{\phi}^2+V}{\dot{\phi}}\right|_{\lambda=H_I^{-1}},
\end{equation}
where  $H$ and $\dot{\phi}$ are evaluated as a given  wavelength $\lambda$
 is stretched beyond
the horizon ($\lambda =H_{I}^{-1}$) during inflation.
  Since all microphysical parameters ($V(\phi)$,
$H$, $\dot{\phi}$, etc.) change slowly during inflation
compared to the stretching rate ($R(t)$),
the ratio in Eq.~(\ref{drho}) is nearly constant for all waves.  That is,
  the fluctuations are
produced with the  (nearly) the same amplitude on average,
a nearly  scale-invariant (Harrison-Zel'dovich \cite{Harr})
spectrum of energy-density  perturbations.

Inflation also generates similar fluctuations in other light fields. For most
fields,
these fluctuations are irrelevant because they are insignificant
contributors to the total energy density, and the fluctuations  leave no
distinctive
signature.   An important exception is
quantum fluctuations of massless gravitons,
%[\Rub, \Sta, \AW],
which result in a nearly scale-invariant spectrum of
gravitational waves \cite{Rub}$^{-}$\cite{AW}.
Because the gravitational waves are weakly coupled, the spectrum is not
erased by reheating or other interactions.  Because of their tensor symmetry,
their signature on the cosmic microwave background anisotropy is
quite different from that of the scalar fluctuations \cite{Dav,Oth}.
The predicted gravitational wave amplitude for a mode
re-entering the horizon is:
   \begin{equation}\label{gamp}
   |h_k|\;\left.\right|_{\lambda=H_I^{-1}} \propto\left.
\frac{H_I^2}{m_p^2}\right|_{\lambda=H_I^{-1}}
 \propto \left. \frac{\frac{1}{2}\dot{\phi}^2+V}
   {m_p^2}\right|_{\lambda=H_I^{-1}},
   \end{equation}
  where $m_p$ is the Planck mass and the expression is to be evaluated
as the wavelength is stretched beyond the Hubble length during inflation.
As with the scalar fluctuations, the parameters in the right-hand expression
change slowly compared to the stretching so that the gravitational wave
spectrum is also nearly scale-invariant.
%BEG2

A critical test for inflation is whether the observed cosmic microwave
background
anisotropy can be explained in terms of the predicted spectrum of scalar and
tensor fluctuations.

\section{What Does Inflation Predict?}

\begin{itemize}
\item {\it Spatial Flatness}:  Inflation
flattens the Universe \cite{Gut}, or, more explicitly,
 exponentially
 suppresses
the spatial curvature contribution to the Hubble expansion relative to
the matter and radiation density and relative to any cosmological
constant ($\Lambda$).
Hence, if $\Omega_{total}$ is defined as  including matter, radiation and
vacuum
energy density contributions, then inflation predicts that $\Omega_{total}=1
\pm \epsilon$, where $\epsilon$ is exponentially small.
\item {\it Gaussian Primordial Perturbations}:  The quantum fluctuations
generated in inflation are Gaussian \cite{G4}.
 For a
gaussian distribution, the total fluctuation spectrum can be determined
from the temperature autocorrelation function (see Section 5).
\item {\it Scale-free Spectrum of Scalar and Tensor Perturbations}:
Energy density and gravitational wave fluctuations are generated during
inflation with a scale-free  spectrum; {\it i.e.}, a spectrum with no
characteristic scale, such as a power-law.
The scalar spectrum at time $t$ is conventionally parameterized in terms of
its Fourier components by a   power-law,
\begin{equation}  \label{nsdef}
\begin{array}{rl}
\left| \left(\left.\delta \rho/\rho\right|_{\lambda=H_I^{-1}} \right)^2\right|
\propto k^{n_s-1} \; &
  \\ &   \\ {\rm or} \;  \; \;
k^3 \left\langle |\frac{\delta \rho}{\rho}(k,t)|^2 \right\rangle
\propto k^{n_s + 3}, &
\end{array}
\end{equation}
 where
$n_s$ is the  called the {\it scalar spectral index}.
The first expression is in terms of $\delta \rho/\rho$ evaluated
at different times,  as each mode
is stretched beyond the horizon during inflation; the second expression
is in terms of $\delta \rho/\rho$ at fixed time.
 In this convention,
$n_s=1$ corresponds to strict scale-invariance (Harrison-Zel'dovich
\cite{Harr}).
The analogous parameterization for the
 gravitational wave spectrum is
\begin{equation}  \label{ntdef}
\left| \left(h_k\;\left.\right|_{\lambda=H_I^{-1}}\right)^2 \right|
\propto k^{n_t}
\; \;  {\rm or} \; \;
k^3 \left\langle |h_{+,\times}(k,t)|^2\right \rangle \propto
k^{n_{t}},
\end{equation}
where $h_{+,\times}$ are the amplitudes of the tensor metric
fluctuations (for two polarizations),
$n_t$ is the {\it tensor spectral index} and
$n_t =0$  corresponds to strict
scale-invariance.\footnote{I apologize in behalf of the   CMB
community for the disgusting
convention that defines the indices such that
$n_s=1$ and $n_t =0$ both correspond to scale-invariant; however,
I will maintain the convention in order for readers of this review
to be able to comprehend the rest of the literature.}

In most inflationary  models, $n_s$ and $n_t$ actually
have  ``weak'' $k$-dependence.   However,
microwave background experiments and large-scale structure measurements
probe only a narrow range of wavenumbers:
Consider a mode with physical wavelength $\lambda$ today.  The
physical wavelength at earlier times is $\lambda'=R \lambda$, where
we choose the convention that $R=1$ today.  The physical wavelength
of the mode at the end of inflation is $\lambda'_{end}=(R_{end}/R_{RH})
R_{RH}\lambda$, where $R_{RH}$ is the value of the scale factor
after the Universe reheats following inflation.  The number of e-folds
of inflation between the time  that the given mode was stretched beyond the
Hubble
distance ($H_I^{-1}$) and the time that  inflation ends is
\begin{equation}
\begin{array}{lcll}
N(\lambda) &  \equiv & {\rm ln}\,\left[\frac{R_{end}}{R_{RH}}R_{RH}\right]
& \\ & & & \\ & = &
57 + {\rm ln} \,\left( \frac{\lambda}{6000 \; {\rm Mpc} }\right) & \\ & & &
\\ &  &
  +\frac{1}{3}
 {\rm ln}\,\left(
\frac{V(\phi_{end})^{1/2}T_{RH}}{ (10^{14}\,GeV)^3}\right), &
\end{array}
\end{equation}
where $V(\phi_{end})$ is the potential energy density at the
end of inflation, $T_{RH}\le V(\phi_{end})^{1/4}$ is the temperature
at reheating, and 6000~Mpc is the present Hubble distance (for $h=.5$).
Microwave background and
large scale structure observations span distances between $\sim 1$~Mpc
and $\sim 6000$~Mpc. These observations cover modes
generated during the 10 e-folds between  $N(1~{\rm Mpc})\sim 50$ and
$N(6000~{\rm Mpc})
\sim 60$.
Over this
narrow range, it is an excellent approximation to
treat $n_s$ and $n_t$ as $k$-independent \cite{Crit92}.
 In the remaining discussion,
$n_s$ and $n_t$ always refer to the values averaged between e-folds 50 and 60.

The total fluctuation spectrum consists of two components, scalar
and tensor, each of which is characterized by an amplitude and a spectral
index.  One convention is   to define the amplitudes in terms of
the scalar and
tensor contributions to the quadrupole moment $C_2^{(S,T)}$
of the CMB temperature
autocorrelation function.  The scalar and tensor fluctuations are
predicted to be statistically independent, so the total quadrupole
is the simply the sum of the two contributions. The
scalar and tensor quadrupole moments are related to the values of
parameters  $N_{H} \equiv N(\sim 6000~{\rm Mpc}) \sim 60$
e-folds before the end of inflation:
\begin{equation}  \label{moms}
 C_{2}^{S} \approx \left. \frac{1}{240 \pi^2}
\frac{H_I^4}{\dot{\phi}^2}\right|_{N_H}
\end{equation}
and
\begin{equation} \label{momt}
 C_{2}^{T} \approx \left. 0.073 \frac{H_I^2}{m_p^2} \right|_{N_H}.
\end{equation}

\item {\it Nearly Scale-Invariant Primordial Spectra}:
The spectral indices are determined by  the equation-of-state, $\gamma$,
  at $50-60$ e-folds before the end of inflation.
The derivation of the relations is straightforward and important, so we
digress to
provide a detailed derivation below.  The reader anxious to
skip to the answers should proceed to
Eqs.~(\ref{nt}) and~(\ref{ns}) and the discussion
below Eq.~(\ref{ns}).

The equation-of-state can be re-expressed
in terms of an inflaton field using the relation:
\begin{equation} \label{gamma}
\gamma =\frac{p}{\rho}=
 \frac{\frac{1}{2}\dot{\phi}^2 - V}{\frac{1}{2}\dot{\phi}^2 + V}=
\frac{8 \pi}{3 m_p^2} \left( \frac{\dot{\phi}^2}{H_I^2}\right) -1,
\end{equation}
where $H_I^2 \equiv (8 \pi/3 m_p^2)[\frac{1}{2} \dot{\phi}^2 +V(\phi)]$.
Instead of $\gamma$, it is useful in this derivation
 to introduce a related parameter:
\begin{equation}\label{alph}
\alpha^2 \equiv 24 \pi \frac{\dot{\phi}^2}{\frac{1}{2}\dot{\phi}^2+V}=
=24 \pi (1+\gamma)=
\left(\frac{8 \pi \dot{\phi}}{m_pH_I }\right)^2.
\end{equation}
A mode that is stretched so that its
wavenumber is $k =H^{-1}$ at  time when there are  $N(\phi)$ e-foldings
remaining
 has wavenumber
$k  = H^{-1} \,{\rm exp}\,N(\phi)$ when inflation ends, where
\begin{equation}
N(\phi) \equiv \int H dt = \int_{\phi}^{\phi_{end}} \frac{H}{\dot{\phi}} d
\phi.\end{equation}
 $N(\phi)$ is the number of e-folds that remain before the
end of inflation  when the
inflaton field has value $\phi$.
Then, we have
\begin{equation}
\frac{d \,{\rm ln}\, k}{d \phi} = \frac{d\, N(\phi)}{d \phi} =
\frac{H}{\dot{\phi}}.
\end{equation}

The tensor fluctuation amplitude as modes are stretched beyond
the Hubble distance inflation is, according to
Eqs.~(\ref{gamp}) and~(\ref{ntdef}), $\propto H_I^2/m_p^2 \propto k^{n_t}$,
Hence, the tensor spectral index can  be computed
according to:
\begin{equation} \label{intnt}
\begin{array}{rlc}
n_t & = & \frac{d \, {\rm ln}\,(H^2/m_p^2)}{d \, {\rm ln}\,k}  \\
& & \\
& = & \frac{d \phi}{d\, {\rm ln}\,k} \frac{d \, {\rm ln}\,(H^2/m_p^2)}{d \phi}
\\
& & \\ & = & \frac{\dot{\phi}}{H^3} (H^2)',
\end{array}
\end{equation}
where the prime will be used to denote derivatives with respect
to $\phi$.
The inflaton satisfies the equation-of-motion
\begin{equation}
\ddot{\phi} + 3 H \dot{\phi}=-V'(\phi),
\end{equation}
but the $\ddot{\phi}$ term is negligibly small during inflation.
This also implies that the kinetic energy density is small
compared to the potential energy density; or,  $(H^2)'\approx 8 \pi V'(\phi)/
3 m_p^2$.
Using the slow-roll equation and the expression for $(H^2)'$, we find that:
\begin{equation}\label{nt}
\begin{array}{rcl}
n_t & =  & -  \frac{8 \pi}{ m_p^2} \left(\frac{\dot{\phi}}{H}\right)^2
 = -\frac{\alpha^2}{8 \pi}\\ & & \\  &  = & -3 (1+\gamma).
\end{array}
\end{equation}
In the end, there is a very simple relation between  $n_t$ and the
equation-of-state.

The analogous derivation for the scalar spectral index is somewhat more
tedious. The amplitude is proportional to $\sim H^4/\dot{\phi}^2 \propto
k^{n_s}$.
Then, we have
\begin{equation}
\begin{array}{rlc}
n_s & = & \frac{d \, {\rm ln}\,(H^4/\dot{\phi}^2)}{d \, {\rm ln}\,k}  \\
& & \\
& = & \frac{d \phi}{d\, {\rm ln}\,k} \frac{d \, {\rm ln}\,(H^4/\dot{\phi}^2)}{d
\phi}
 \\
& & \\
& = & 2 \frac{\dot{\phi}^2}{H^3}\left[\frac{2 H H'}{\dot{\phi}} -
\frac{H^2}{\dot{\phi}^2}\dot{\phi}'\right] \\
& & \\
& = &  \frac{\dot{\phi}}{H^3} (H^2)' +2 \left(\frac{\dot{\phi}
H'-\dot{\phi}'H}{H^2}\right).
\end{array}
\end{equation}
The first term,  precisely the same
as the intermediate expression we obtained for $n_t$ in Eq.~(\ref{intnt}),
is $-3(1+\gamma)$.  By the use of
 Eq.~(\ref{alph}) and a bit of algebra, the second
term can be expressed as
 $$- \frac{m_p}{4 \pi} \frac{d \alpha}{d\phi}=
\frac{d\, {\rm ln}\,\alpha^2}{
d\, {\rm ln}\, k} = \frac{d\, {\rm ln}\,(1+\gamma)}{d\, {\rm ln}\, k}.$$
Hence, the total expression is
   \begin{equation}\label{ns}
   n_s =
1- 3 (1+\gamma) +
   [d \, {\rm ln} \; (1 + \gamma)/d\,{\rm ln}\,k];
   \end{equation}

Strict exponential (de Sitter) expansion,
$R(t) \propto {\rm exp}(H_I t)$, corresponds to  $\gamma=-1$, in which
limit one obtains precise scale-invariance, $n_s=1$ and $n_t=0$,
according to Eqs.~(\ref{ns}) and~(\ref{nt}).
However, in any realistic inflation model, the expansion rate must
slow down near the end of inflation in order
to return to  Friedmann-Robertson-Walker expansion.  If $R(t)$ is
inflating but
not exponentially, then
$-1/3>\gamma>-1$,  $\gamma'$ may not be zero,  and
$n_s \ne 1$ and $n_t<0$.   Inflationary models fall in the range
$0.7 \lesssim n_s \lesssim 1.2$ and $0.3 \lesssim n_t \le 0$; pushing
$n_s$ and $n_t$ beyond this range entails exceptional models
with special choices of
parameters and/or initial conditions \cite{Crit92,Hodge}.
 (See comments at the end
of this section.)
\item {\it Relations between $(C_2^{(S)},\; C_2^{(T)},\; n_t, \;
n_t)$}:
COBE and other large-angular scale experiments can determine the
total quadrupole moment, $C_2 = C_2^{(S)}+ C_2^{(T)}$, placing
one constraint on the four parameters that define the inflationary
spectrum.
  The three remaining degrees of
freedom are: $r \equiv C_2^{(T)}/C_2^{(S)}$, $n_s$ and $n_t$.
These three parameters are all expressible in terms of the
equation-of-state, $\gamma$, and, hence, can be related to one
another \cite{Dav,Oth}.  The tensor-to-scalar quadrupole ratio, $r$,
is obtained by taking the ratio of Eq.~(\ref{momt}) to
Eq.~(\ref{moms}): $r \approx 173 \dot{\phi}^2/(H_I^2 m_p^2)$. Using
Eq.~(\ref{gamma}), we can convert this to:
\begin{equation}
r \equiv C_2^{(T)}/C_2^{(S)} \approx  21 (1+\gamma).
\end{equation}

Comparing the last relation  to Eq.~(\ref{nt}), we find that
\begin{equation} \label{rel1}
r\approx -7  n_t.
\end{equation}
or, we can compare it  to Eq.~(\ref{ns}) and obtain
\begin{equation}\label{rel2}
   n_t=  n_s -1 -
 [d \, {\rm ln} \; (1 + \gamma)/d\,{\rm ln}\,k],
\end{equation}
and
\begin{equation}\label{rel3}
r \approx 7(1- n_s)+
[d \,{\rm ln} \; (1 + \gamma)/d\,{\rm ln}\,k]
\end{equation}
(N.B. $r$  is non-negative, by definition. It is possible to
construct models in which the right-hand-side
of Eq.~(\ref{rel3}) is formally negative; this result should
be interpreted as indicating negligible tensor contribution, $r\approx 0$.
In particular,
models with $n_s>1$, such as some hybrid
inflation \cite{hyb} models, have negligible tensor fluctuations.)
These three relations constitute a set of  testable
 signatures of inflation.   If observations establish
that the primordial spectrum of perturbations is scale-free,
observational support for  these additional relations would be
evidence
 that the perturbations were generated by inflation.
\end{itemize}

\begin{figure}
%\centerline{\psfig{file=snowf1a.ps,width=3.3in}}
\caption{
The range of the $r$-$n_s$ plane consistent with  generic
inflationary models is
enclosed by the box.  Most models are constrained to lie along the grey
 diagonal
curve;  models in which the inflaton encounters an
extremum in the inflaton potential near the last
60 e-folds have negligible tensor contribution, $r\approx0$,
along
the abscissa inside the box.}
\label{fig:f1}
      \end{figure}

The predictions are partially  illustrated in Fig.~1, which shows a range of
parameter-space in the $r$-$n_s$ plane.   Whereas the entire plane
describes tensor and scalar perturbations which are scale-free, the
range allowed by inflation
is confined to the range  $0.7 \lesssim n_s \lesssim 1.2$, which is nearly
scale-invariant.
Then, Eq.~(\ref{rel3}) places a constraint on $r$.  Hence, over the
entire $r$-$n_s$ plane, inflationary predictions are confined to a small box.
(The  boundaries of the box are not precisely defined; one
can expand the box 10 per cent or so at the cost of additional fine-tuning
of parameters.)

Inflation is falsifiable if observations show
that the CMB spectrum lies far outside the box.  For example, early
reports from COBE analysis
suggested that $n_s > 1.5$ \cite{Wright}, which would be
 inconsistent. These results have since
been revised to $n_s \approx 1.0 \pm 0.3$, which is consistent with
inflation \cite{Gorski,Bondn}.

Fig.~1 further illustrates that the predictions of
inflation do not uniformly cover the box.  For most   models,
$d\,{\rm ln}\,(1+\gamma)/d\,{\rm ln}\,k$
is negligible during inflation (which is a way of saying
that the equation-of-state changes
very slowly during inflation), and, hence,
\begin{equation} \label{rnsapprox}
r\approx 7(1-n_s).
\end{equation}
These models lie along
the grey curve   of negative slope shown within the box.
A   subclass of models has the property that the inflaton encounters
an extremum of the inflation potential 60 or so e-folds before the
end of inflation.
In these models,
 $d\,{\rm ln}\,(1+\gamma)/d\,{\rm ln}\,k
\propto V''(\phi)$ changes significantly
during inflation; in particular, $\dot{\phi} \propto V'(\phi)$
 shrinks significantly near
the extremum, which   amplifies the scalar perturbations
relative to the tensor (see Eqs.~(\ref{drho}) and~(\ref{gamp})).
Consequently, these models predict that $r\approx 0$, along the abscissa of
Fig.~1.
 Evidence that the CMB spectrum lies in the box but
far from the abscissa or the ($r\approx 7(1-n_s)$) diagonal would
be problematic for inflation.

The ``generic" predictions of inflation outlined above presume
no theoretical prejudice about the ``brand" of inflation.
They apply to  new, chaotic,
supersymmetric, extended, hyperextended, hybrid and natural inflation.
What determines the prediction is the equation-of-state ({\it e.g.},
 the shape of the inflaton
potential $V(\phi)$) during
the last 60 e-folds of inflation.
Table 1 summarizes the predictions of inflation for some particular
forms of the inflaton potential, $V(\phi)$.
Since the examples in the Table
run the gamut from potentials which are steep to those which are
flat, it may be used to estimate the predictions
for more
general potentials.   Conversely, it is possible to reconstruct
a section  of the inflationary
potential over the
  range of $\phi$
covered during e-folds 50 to 60 from measurements of the
spectral index and the ratio of tensor-to-scalar quadrupole
moments \cite{recon}.  This reconstructed section
is extremely narrow since $\phi$ evolves only  a tiny distance
down the potential during e-folds 50 to 60.  Furthermore,
the reconstructed section typically lies far from the false or
true vacuum.  Hence, the reconstructed section is of limited value in
determining the full potential or underlying physics driving inflation.

\begin{table}[t]
\begin{center}
\begin{tabular}{||c|c|c|c||}
\hline \hline & & & \\
$V(\phi)$ & $n_s-1$ & $n_t$ & $r$ \\
& & & \\
\hline \hline & & & \\
$V_0 {\rm exp}(-\frac{c\phi}{m_p})$ &$ -\frac{c^2}{8 \pi}$ &
$- \frac{c^2}{8 \pi}$ & .$28 c^2$ \\ & & & \\  \hline & & & \\
$A \phi^n$ & $-.02 - \frac{n}{100}$  & $-\frac{n}{100}$
& $.08 n$ \\ & & & \\  \hline  & & & \\
$V_0 + \lambda \phi^4 ({\rm ln}\,\frac{\phi^2}{\sigma^2} \, -\,\frac{1}{2})$
 &
$-4 \times 10^{-6}\left(\frac{\sigma}{m_p}\right)^4$ &
$-.06 -4 \times 10^{-6}\left(\frac{\sigma}{m_p}\right)^4$ &
$3 \times 10^{-5} \left(\frac{\sigma}{m_p}\right)^4$ \\   & & & \\ \hline
& & & \\
$V_0 \left(1-\frac{\phi^2}{f^2}\right)$ & $- \frac{m_p^2}{2\pi f^2}$
& $-\frac{\pi m_p^2}{8 f^2} {\rm e}^{-N m_p^2/2 \pi f^2}$
&$ 2.8 \frac{m_p^2}{f^2} {\rm e}^{-N m_p^2/2 \pi f^2}$\\ & & & \\
\hline \hline
\end{tabular}
\vspace{0.5cm}\end{center}
\caption{
Predictions of inflationary models for some common potentials.
}
\label{tab:infla}\vspace{0.5cm}]\vspace{-1.5cm}
\end{table}

In the Table,
 $m_p\equiv 1.2 \times 10^{19}$~GeV is the Planck mass, $N=60$
is e-folds corresponding to the present horizon ({\it i.e.},
the natural log of the
 ratio of the present horizon to the horizon during inflation
in comoving coordinates).
The first two examples correspond to cases where Eq~{\ref{rnsapprox}}
is a good approximation. The approximation correctly predicts a negligibly
small value of $r$ for the third example, but the   numerical
value is not well-estimated. The predictions for these first three
models lie close to the diagonal line in Fig.~1.  Eq~{\ref{rnsapprox}}
is a poor approximation for the fourth
example, in which the inflaton rolls from the top of a quadratic
potential and $d\,{\rm ln}\,(1+\gamma)/d\,{\rm ln}\,k$ is large.  The
prediction for this case is
a spectrum with
 tilt ($n_s<1$) but insignificant gravitational wave contribution,
corresponding to points along the abscissa of Fig.~1.

Exceptional inflationary
models can be constructed which violate any or all of the conditions
described above.   In fact, because theorists enjoy dwelling on
such matters, there are nearly as many papers written on exceptional
models as on generic ones.  This causes some experimentalists,
observers, and non-experts to give these exceptional predictions
undue weight.   Therefore, it is important to emphasize that
these exceptional models, such as those which predict
an open or closed Universe or
primordial spectra which are not scale-free,
 are extremely unattractive.
First, they require extraordinary fine-tuning
beyond what is required to have sufficient inflation and solve
the conundra of the big bang model.  If one maps out
the range of parameter-space which gives sufficient inflation,
the exceptional models occupy an exponentially tiny corner.
Second, the predictions are not robust.   Moving from one point
to another within the tiny corner of parameter-space
significantly changes the predictions.  For example, if one
choice of parameters produces a Universe with $\Omega=0.1$,
a slightly different choice increases or decreases the number
of e-foldings by one, which results in a change in $\Omega$ by
a factor of ten.    Consequently, there is little or no
real predictive power to the exceptional models.

Focusing on the generic tests of inflation  is well-motivated
for broader reasons:
  The same tests  might determine whether
the Universe can be explained on the basis of physical laws
testable in the laboratory ({\it Path I}, as defined in Section 2).
The microphysics which we presently understand or can hope to
test in the laboratory involves time-scales
 infinitesimally smaller than the age of the Universe and length-scales
infinitesimally smaller than the Hubble length (or the sizes of galaxies).
If our Universe is to be  comprehended from these
physical laws alone without
special choices of initial conditions or parameters ({\it i.e.}, {\it Path I}),
we should not find that
there is something special about
the present epoch (compared, say, to 10 Hubble times from now or 10 Hubble
times earlier) or that  there are special features (bumps, dips, etc.) in the
primordial spectrum of fluctuations of cosmological wavelength.
If we find evidence for new time- or length-scales of cosmological
dimensions which cannot be probed in the laboratory, cosmology is thrust
into {\it Path II}.   Important features of our Universe must be
attributed to initial conditions or physical laws which
probably  can never be independently tested.
(N.B.  Testing whether the spectrum is scale-free
  is less specific than testing
inflation.  One can imagine finding evidence which supports
a flat Universe with scale-free primordial perturbations, but
which conflicts with
the inflationary relations between $r$ and $n_s$.)

\section{Translating Inflationary Predictions into Precision Tests}

The predictions of inflation described in the previous Section
can be translated into
 precise tests of the CMB
anisotropy.  The implications for the CMB anisotropy
can be obtained by  numerical integration of the general
relativistic Boltzmann, Einstein, and hydrodynamic
equations \cite{Crit93}.
Included in the dynamical evolution are all the
relevant matter-energy
components: baryons, photons, dark matter, and massless
neutrinos.   The temperature anisotropy, $\Delta T/T\, ({\bf x})=
\sum_{\ell m} a_{\ell m} Y_{\ell m} (\theta,\phi)$, is computed in
terms of scalar \cite{Efst,BondE}
 and tensor \cite{Crit93} multipole components, $a_{\ell m}^{(S)}$
and $a_{\ell m}^{(T)}$, respectively.

The common method of characterizing the CMB fluctuation spectrum
is in terms of multipole moments. Suppose that one measures
the temperature distribution on the sky, $\Delta T/T\, ({\bf x}$).
The temperature
autocorrelation function (which compares points in the sky
separated by angle $\alpha$)  is defined as:
\begin{equation} \label{autoc}
\begin{array}{rcl}
C(\alpha) &=&
\left\langle{\frac{\Delta T}{T}({\bf x}) \frac{\Delta T}{T}({\bf x'}) }
\right\rangle  \\ & & \\ &
= & \frac{1}{4 \pi}
\sum _{\ell} (2 {\ell}+1) C_{\ell} P_{\ell}({\rm cos}\; \alpha),
\end{array}
\end{equation}
where $\left\langle\right\rangle$ represents  an average over the
sky and ${\bf x}\cdot{\bf x'}= \;  {\rm cos}\; \alpha.$
The coefficients, $C_{\ell}$, are the {\it multipole moments}
(for example, $C_2$ is the quadrupole, $C_3$ is the octopole, {\it etc.}).
Roughly speaking,  the value of
$C_{\ell}$ is determined by fluctuations on angular scales $\theta
\sim \pi/\ell$.
A plot of $\ell (\ell+1) C_{\ell}$ is referred to as the {\it power
spectrum}.
This definition is chosen so that an exactly scale-invariant
spectrum, assuming no evolution when the fluctuations pass inside
the Hubble horizon, produces a flat power spectrum ({\it i.e.},
 $\ell (\ell+1) C_{\ell}$ is independent of $\ell$).
  For inflation,  the contributions of scalar and tensor fluctuations
to the $a_{\ell m}$'s  are
  predicted to be statistically independent.
Consequently, the total  multipole moment $C_{\ell}$
is the sum of the scalar and tensor contributions, $C_\ell^{(S)}$
and $ C_\ell^{(T)}$, respectively. The  $a_{\ell m}$'s are also
predicted to be Gaussian-distributed.  The cosmic mean value predicted by
inflation is
$C_\ell^{(S)}= \left\langle \vert a_{\ell m}^{(S)}\vert^2
\right\rangle_E
$ and
$ C_\ell^{(T)}= \left\langle \vert a_{\ell m}^{(T)}\vert^2
\right \rangle_E$ where these  are averaged over an ensemble of universes
$\{E\}$
and over $m$.  The $C_{\ell}$'s, which are an average over $2\ell+1$
 Gaussian-distributed variables, have a  $\chi^2$-distribution.

There is additional valuable
information in higher-point temperature correlation functions; {\it e.g.},
tests for non-gaussianity.  However,  statistical and systematic
errors increase for higher-point correlations; for the
short run, the most reliable information will be
 the  angular power spectrum ($ C_{\ell}$ vs. $\ell$).

The predicted spectrum depends not only on the inflationary
parameters ($r$, $n_s$, $n_t$) but also on other cosmological parameters.
Inflation
produces a flat Universe, $\Omega_{total} \approx 1$.  In the examples
shown in this article, we assume no hot dark matter,
$\Omega_{HDM} =0$, but note that, for angular scales $\gtrsim
10^\prime$, the anisotropy for mixed hot and cold dark matter models with
$\Omega_{CDM} + \Omega_{HDM} \approx 1$ is quite similar to the
anisotropy if all of the dark matter is cold. For a given value of the
Hubble parameter, $H=100 h$~km/sec/Mpc, we
impose the nucleosynthesis estimate,
%[\bbn],
$\Omega_B  h^2 =
0.0125$, to determine $\Omega_B$. We also satisfy globular cluster
 and
other age bounds \cite{hub},
%[\kolb],
and gravitational lens limits \cite{grav}:
%[\eturner]:
we
range from $h\lesssim 0.65$ for $\Omega_{\Lambda}=0$ to $ h
\lesssim 0.88$ for $\Omega_{\Lambda} \lesssim 0.6$.
We also consider a range of reionization scenarios in which the intergalactic
medium is fully reionized at some red shift $z_R$ after recombination.

\begin{figure}
%\centerline{\psfig{file=snowf2a.ps,width=3.35in}}
\caption{Predicted power spectrum predicted for inflationary
cosmology.
The spectra are for
 h=0.5,
$\Lambda=0$,  $\Omega_B h^2 =.0125$ and cold dark matter (CDM).
The upper (solid)  curve has spectral index $n=1$
(Harrison-Zel'dovich) and pure
scalar  fluctuations, $r=0$;  the vertical hashmarks represent
the one-sigma, full-sky cosmic variance.
The lower (dashed)  curve  has
 $n=0.85$ and  50-50 mixture of scalar and tensor
quadrupole perturbations, $r=1$.
}
\label{fig:f2}
\end{figure}

Fig.~2 shows the predicted power
spectrum
for the  central range of inflationary
models consistent with the generic predictions outlined in Section 4.
Since the value of $C_{\ell}$ is determined by fluctuations
on angular scales $\theta\sim \pi/\ell$, moving left-to-right
in the Figure corresponds to moving from large-angular scales to
small-angular scales.  Since large-angular scale fluctuations entered the
Hubble horizon recently compared to small-angular scale fluctuations,
moving left-to-right also corresponds to moving from unevolved primordial
fluctuations to fluctuations which have evolved inside the Hubble horizon
for a significant time.
For these examples, we have chosen $h=0.5$, $\Omega_{\Lambda}=0$, and
 $\Omega_B =0.05$.

The predictions of theoretical models, including those for inflation
shown in Fig.~2, are expressed in terms of
the cosmic mean value of the $C_{\ell}$'s
averaged over an ensemble of universes or, equivalently,
over an ensemble of Hubble horizon-sized patches.   In reality,
experiments can  measure, at best,  over
a single Hubble horizon distance.  Since the experiments are
limited in this way, it is important to know not only the
theoretical prediction for the cosmic mean, but also the
the theoretical variance about  that prediction for experiments
 confined to a single Hubble horizon distance.   This  uncertainty,
known as the full-sky ``cosmic variance,"
 is equal to $C_{\ell}/\sqrt{2 \ell+1}$
for Gaussian-distributed fluctuations, such as those predicted by
inflation.  Note that the variance decreases
with increasing $\ell$.
Many
current small-angular scale experiments
 cover only  a tiny fraction of Hubble horizon-sized patch.  If the
area fraction of full-sky coverage is $A$,
the theoretical uncertainty
scales as $A^{-\frac{1}{2}}$.
(In cases where there
is much less than full-sky coverage, the theoretical uncertainty
 is often referred to as ``sample
variance.")

A more realistic situation is where there are errors due to
both sample variance
and detector noise \cite{Bond94}.
Consider a detection obtained from
measurements $(\overline{\Delta T/T})_i \pm \sigma_D$
($\sigma_D$ represents detector noise) at $i=1,\ldots
,N_D$ experimental patches sufficiently isolated from each other
to be largely uncorrelated.
 For large $N_D$, the likelihood function falls by
$e^{-\nu^2 /2}$ from a maximum at $(\Delta T/T)_{max}$ when
\begin{equation}  \label{var}
\left(\frac{\Delta T}{T}\right)^2 = \left(\frac{\Delta T}{T}\right)_{max}^2
 \pm \sqrt{\frac{2}{N_D}}\, \nu \,  [\left(\frac{\Delta T}{T}\right)_{max}^2 +
    \sigma_D^2] \, .
\end{equation}
%\begin{equation}  \label{var}
%\left(\Delta T/T\right)^2 = \left(\Delta T/T\right)_{max}^2
% \pm \nu \sqrt{(N_D/2)}\,  [\left(\Delta T/T\right)_{max}^2 +
%    \sigma_D^2]\, .
%\end{equation}
An experimental noise $\sigma_D$ below $10^{-5}$ is standard now, and
a few times $10^{-6}$ will soon be  achievable; hence, if systematic errors
and unwanted signals can be eliminated, the one-sigma ($\nu =1$) relative
uncertainty in $\Delta T/T$ will be from cosmic (or sample) variance alone,
$1/\sqrt{2N_D}$, falling below $10\%$ for $N_D > 50$.
The hashing in Fig.~2 corresponds to full-sky cosmic variance,
roughly equivalent  to filling the sky with
patches separated by $2\theta_{fwhm}$.

%FinalChanges1: Begin here
\section{The Fingerprint of Inflation}

The CMB power-spectrum (Fig.~2) is a fingerprint which
is evidence both for inflation and for certain values of
cosmological parameters.
 $C_{\ell}$'s for
$\ell \lesssim 200$ are dominantly determined by fluctuations outside
 the Hubble horizon at recombination.   These multipoles
 measure  the primordial spectrum of fluctuations.
 $C_{\ell}$'s for
$\ell \gtrsim 200$ are sensitive to fluctuations inside the Hubble
horizon at recombination.  These fluctuations, which
 had time to evolve prior to
 last scattering,
are sensitive to evolutionary effects which depend on
 the matter density, the expansion rate,  and
the density, type, and distribution of dark matter.
Fig.~3 shows a schematic of the spectrum and the various
contributions to it.
The distinctive features of the fingerprint are
(reading Fig.~2 from left to right):

\begin{figure}
%\centerline{\psfig{file=waynehu.ps,width=4.in}}
\caption{
Schematic of the power spectrum predicted by a typical
inflationary model and the major contributions to it:
the Sachs-Wolfe effect (scalar and tensor modes combined),
acoustic oscillations of the
baryon-photon fluid (density and velocity contributions),
and the integrated Sachs-Wolfe effect
(significant in models where recombination occurs before
matter-domination). [Adapted from Hu \& Sugiyama 1994.]
}
\label{fig:wayne}
\end{figure}

\noindent
{\it Plateau at large angular scales ($\ell\lesssim 100$)}
 is due to fluctuations in the gravitational
potential on the last-scattering surface (the Sachs-Wolfe effect\cite{SW}).
Fluctuations in the potential induce red shifts and blue shifts in the
CMB photon distribution which create apparent temperature fluctuations.
For a precisely scale-invariant  ($n_s=1$) spectrum of scalar fluctuations,
the Sachs-Wolfe contribution to
$C_{\ell}$ is proportional to $1/(\ell(\ell+1))$, and so the
contribution to  the
power spectrum,  $\ell (\ell+1)C_{\ell}/2 \pi$ (the ordinate in Fig.~2),
is flat.
  The full computation reveals a slightly upward slope at small $\ell$
due to other, higher-order contributions, the same effects which
are responsible for the Doppler peaks described below.
  For $n_s<1$, there is a downward tilt to the Sachs-Wolfe
contribution relative to $n_s=1$ (less power
on smaller scales).  The $n_s<1$ curves in Fig.~2 include the tensor
contribution predicted by generic inflation models, as
expressed by Eq.~(23).
{}From observations at small $\ell$ only, it is difficult to distinguish the
 tensor contribution because
the spectral slope due to  the Sachs-Wolfe  effect is nearly the same
for tensor and scalar contributions for $\ell>3$.\footnote{There is an
intriguing,  notable
difference in the ratio of the mean  quadrupole-to-octopole moment for the
models with tensor contribution. However,  the effect is not a useful
discriminant because there is a
large  cosmic variance for the small-$\ell$  multipoles.}
The slope is not very
 sensitive to the value of $h$, $\Omega_B$, $\Omega_\Lambda$
or other cosmological parameters.

\noindent
{\it First Doppler Peak at $\ell\approx 200$} probes
wavelengths smaller than the horizon at last-scattering ($\lesssim 1^{\circ}$).
Gravitational wave perturbations begin to oscillate and red shift
away once their wavelength falls within the horizon.
Hence, for $\ell \gtrsim 200$,
the tensor perturbations do not contribute significantly,
 even if they were dominant over the scalar fluctuations
at larger angular scales ($\ell \ll 200$).

The prominent features,
   known as the Doppler peaks, are due to scalar fluctuations.  Fig.~3
illustrates
 both the acoustic density and acoustic velocity contributions.
The peaks are the remnant of adiabatic oscillations in the baryon-photon fluid
density.
The oscillations in a given mode begin
when the wavelength falls below the Jeans length ({\it i.e.}, pressure
dominates
over gravity).\footnote{The term, ``Doppler peak," is a misnomer since, with
standard recombination, the
electrons and photons oscillate together and there is little difference
in their velocities.
The Doppler effect does not become significant
until  $\ell\gtrsim 1000$.} The Jean's length near
recombination  is
roughly  $2 \pi c_s H^{-1}$, where $c_s \approx 1/\sqrt{3}$
is the sound speed.

The value of $\ell$ at the maximum \cite{kamion}
of the first Doppler peak
 is
 $\ell_{peak} \approx 220/
\sqrt{\Omega_{total}}$.   Since the location
is insensitive to the value of $h$ or $\Omega_B$, measuring $\ell_{peak}$
is a novel means of measuring $\Omega_{total}$.
The height of the Doppler peak depends on the primordial spectral amplitude,
the scalar spectral index ($n_s$),  the tensor-to-scalar quadrupole
ratio ($r$),
the expansion rate, and the pressure \cite{Bond94}.
If the spectrum is fixed at large angular scales by COBE DMR,
smaller values of $n_s$ imply decreasing primordial amplitudes
 on smaller angular scales and,
consequently, a smaller Doppler peak.
Gravitational waves add to
the plateau at $\ell < 200$,  but their contribution to the Doppler
peak is red shifted to insignificant values.
Consequently,  increasing $r$ decreases the height of the Doppler
peak relative to the plateau.
Increasing the expansion rate ($h$)
pushes back matter-radiation equality relative
to recombination, thereby
increasing the adiabatic growth of perturbations.  Photons escaping from the
deeper gravitational potential  are red shifted more.  Hence, increasing $h$
 suppresses the Doppler peak.  Increasing  the pressure (by decreasing
$\Omega_B h^2$)  also decreases the anisotropy
since the fluctuations stop growing once pressure dominates the
gravitational infall.  (According to the last two
remarks, increasing $h$ produces opposite
effects.
 For $\Omega \lesssim 0.1$,
the net effect is  that  increasing $h$ decreases the anisotropy \cite{hu}.)
The height of the first Doppler peak is relatively insensitive to
whether the dark  matter is cold or a mixture of hot and cold dark matter.

\noindent
{\it Second and Subsequent Doppler peaks}
are due  to modes
that have undergone  further  adiabatic oscillations.
The peaks are roughly periodically-spaced.
The deviation from periodicity is due
to time-variation in $c_s$.
The amplitudes
decrease as $n_s$ decreases and $r$ increases.

Anisotropies are caused by inhomogeneities in the baryon-photon
fluid density and velocity.  The acoustic
density and velocity oscillations are 180 degrees
out-of-phase with one another, as shown in Fig.~3.
The acoustic density contribution
is larger.  The Doppler peak maxima and minima correspond to  maxima  and
minima of
the acoustic density oscillations;  the minima do not extend to zero
because they are
  filled in by the maxima in the acoustic
velocity contribution \cite{BondE,hu,seljak}.
The first and other odd-numbered peaks correspond to compressions and the
even-numbered peaks correspond to rarefactions.  Gravity tends to enhance
the compressions and suppress the rarefactions.  The effect is especially
noticeable at low pressure (high $\Omega_B h^2$), for which the even-numbered
peaks are greatly suppressed or absent altogether.
The  peaks are  also
 sensitive to whether the dark matter is cold or a
mixture of hot and cold.
%FinalChanges1: End Here

{\it Damping at $\ell \gtrsim 1000$}:
CMB fluctuations are suppressed by photon diffusion (Silk Damping \cite{silk}).
The baryons and photons are imperfectly coupled. The photons
tend to diffuse out of the fluctuations and smooth their distribution.
Through their collisions with the baryons, the baryons
distribution is smoothed as well,
 thereby suppressing the anisotropy.
A second damping effect is due to the destructive interference of
modes with wavelengths smaller than the thickness of the last-scattering
surface \cite{Efst,BondE}.  Doppler peaks from five or so adiabatic
oscillations can
be distinguished before the damping overwhelms.

\section{Five Milestones for Testing Inflation}

The fingerprint imprinted by inflation  on the CMB anisotropy suggests
a series of five milestone tests.
Below, the proposed tests are compared
with current experimental
results.  Included are \cite{cobe}$^{-}$\cite{ovro}:
%\cite{cobe,firs,ten,sp,bp,python,argo,max,msam,wd,ovro}
 COBE DMR (COsmic Background Explorer
Differential Microwave Radiometer), FIRS (Far InfraRed Survey),
TENerife, SP91 and SP94 (South Pole 1991 and 1994),
BP (Big Plate), PYTHON, ARGO,
MAX (Millimeter Anisotropy
Experiment), MSAM (Medium Scale Anisotropy
Experiment), White Dish,
and OVRO7 (Owens Valley Radio Observatory, 7 degree experiment).

\noindent
{\bf Milestone 1: Observation of Large-Scale Fluctuations with
$\Delta T/T \approx 10^{-5}$}

The nearly scale-invariant spectrum of fluctuations generated
by inflation includes modes whose wavelength is much greater
than the Hubble horizon at recombination ($\gg 1^{\circ}$).
If inflation is  responsible for the formation of
large-scale structure,
the magnitude of the perturbations should be at the
level of $\Delta T/T \approx 10^{-5}$.   If $\Delta T/T$ were
a factor of five or more smaller,  the amplitude would be
too small to account for galaxy formation.
A value of  $\Delta T/T$ a factor of
five or more greater  would lead to unacceptable clumping of
large-scale structure.

In starting a long journey, it is encouraging  to begin from a point
where the first milestone has already been passed.
In this case, COBE DMR \cite{cobe},
with some corroboration from the FIRS \cite{firs} and Tenerife \cite{ten}
 observations,
 finds $\Delta T/T = 1.1 \pm 0.1 \; \times \;10^{-5}$ scales (two-year
result for  53 GHz scan smoothed over $10^{\circ}$) \cite{Wright}, just
within the range consistent with inflation and dark matter
models.  Other models, such as cosmic defects and isocurvature
baryon (PIB) models, are also consistent with these observations.

\vspace{.1in}
\noindent
{\bf Milestone 2: Observation of Scale-free and Nearly Scale-Invariant
  Spectrum at Intermediate Angular Scales ($20^{\circ} \gtrsim \theta
\gtrsim 1^{\circ}$)}

Inflation predicts a primordial
 spectrum  that
is scale-free and nearly scale-invariant.  Fig.~4 shows the
predictions
for the central range of parameters
consistent with inflation  compared with current measurements of
the
 CMB anisotropy.
The curves have been normalized to the COBE DMR two-year
result.

A notable effect is that the spectrum for the $n_s=1$,
scale-invariant (Harrison-Zel'dovich) spectrum is not precisely flat.
The slight, upward tilt is due to the small contributions of short
wavelength ($\lesssim 1^{\circ}$) modes, which include effects
other than simple Sachs-Wolfe.  The same contributions
become dominant  at
$\ell \ge 100$ and are responsible for the Doppler peaks addressed
by Milestones 3 and~4.
Consequently, the  {\it apparent}
spectral index  --- the slope of $C_{\ell}$ vs.~$\ell$
determined directly from the CMB anisotropy --- differs from  the
{\it primordial} spectral index which generated the spectrum.
For example, the upper curve has been computed for a primordial
index of $n_s=1$, but that apparent index (the upward tilt)
corresponds to $n_s^{app}=1.15$, a 15 per cent correction.
An important feature of the added contributions
 is that, over the range
$\ell \lesssim 100$, they are relatively insensitive to the value of
cosmological parameters ($h$, $\Omega_B$, and $\Omega_{\Lambda}$).
Fig.~6 illustrates the range of predictions for fixed
spectral index ($n_s=1$)
 when all other parameters are varied by the maximal
amounts consistent with astrophysical observations.
The spectra form a narrow sheath around the original
spectrum, clearly separated from the prediction for $n_s=0.85$.
Hence, the spectrum of $C_{\ell}$'s for $\ell \lesssim 100$ can
test whether the spectrum is scale-free and measure the spectral
index without any additional assumptions about other cosmological
parameters.

Fig.~4 also shows current detections. The error bars
represent the one-sigma limits.
(The method of flat band power estimation \cite{Bondonly,capri},
an important tool for converting
experimental results into model-independent
bounds on the  $C_{\ell}$'s,  is discussed in
the Appendix.)
The theoretical curves are normalized
to match COBE DMR (two-year).
The experiments on this figure
illustrate the diversity
in CMB experiment.  COBE DMR is aboard a space-borne platform;
FIRS is a high-altitude balloon experiment;  Tenerife is
mounted on a mountain in the Canary Islands; South Pole is in
Antarctica, of course; and Big Plate is set on the ground in Saskatoon.

At this point, experiments are consistent with a
scale-free form for the power spectrum with spectral-index
near $n_s=1$ and $n_t=0$, but the error bars are large.
After four years of data, COBE DMR may be able to reduce
its error by two. Much more dramatic improvements are
expected for the other experiments.   The
chief limitation is small sample area (and, hence,
overwhelming sample variance) which should  be overcome with
continued measurements.

By itself, COBE DMR is not a powerful discriminant among theoretical
models. For example, Fig.~7 shows the cross-correlation
between the 53~GHz and 90~GHz frequency channels.  The cross-correlation
for a full-sky map is
$K(\alpha) \equiv
\langle \Delta T_{53}({\bf x}) \Delta T_{90}({\bf x}')\rangle$ where
${\bf x}\cdot{\bf x}' = {\rm cos} \;\alpha$;
the coefficients of the multipole expansion of $K(\alpha)$
analogous to  Eq.~(\ref{autoc}) are $K_{\ell}$.  If COBE DMR provided
a full-sky map and there was no detector noise or foreground contamination,
the $K_{\ell}$ should equal $C_{\ell}$.  In reality, COBE DMR analysis
is based on a full-sky from which the galaxy has been cut.
Plotted in Fig.~7 are the coefficients of orthonormal
moments on the cut sky, using methods developed by Gorski \cite{Gorski}.
The fact that there is noise in the COBE
experiment accounts for the anticorrelation at large $\ell$.
At $\ell \lesssim 20$, though, the two channels are highly correlated
(open circles)  and there is strong evidence for the CMB signal.
Superimposed on the plot are predictions for the cross-correlation for
 diverse models. The dotted lines correspond to one-sigma cosmic
variance.  The figure shows that the models cannot be distinguished
to better than cosmic variance.

%66 lines
\begin{figure}
%\centerline{\psfig{file=snowf3.ps,width=4.0in}}
\vspace{0.0cm}
\caption{
Blow-up of Fig.~2 showing the power spectrum for intermediate
angular scales, $2 < \ell<100$. Note that the upper curve, which
corresponds to $n_s=1$ (scale-invariant), has an upward
tilt corresponding to an apparent spectral index of $\sim 1.15$
at large angular scales.
Superimposed are the experimental flat band power detections
with one-sigma error bars
(see Appendix)
%FinalChanges2: Begin Here
for: (a) COBE; (b) FIRS; (c) Tenerife;
(d) South Pole 1991; (e) South Pole 1994;  (f)
Big Plate
1993-4;
and (g) PYTHON.
}
%FinalChanges2: End Here
\vspace{0.5cm}
%\vspace{-1.45cm}
\vspace{-1.25cm}
\label{fig:f3}
 \end{figure}

\begin{figure}
%\centerline{\psfig{file=snowf5.ps,width=4.0in}}
\vspace{0.0cm}
\caption{
CMB spectra through the first Doppler
peak.
The solid and dashed black curves correspond
to the inflationary/CDM predictions for spectral index
($r=0$, $n_s=1$) (Harrison-Zel'dovich) and ($r=1$, $n=0.85$),
respectively.  The  dot-dashed curve is the prediction for
cosmic texture models.
For the data, error bars represent one-sigma.
%FinalChanges3: Begin Here
The
experiments correspond to: (a) COBE; (b) FIRS; (c) Tenerife;
(d) South Pole 1991; (e) South Pole 1994; (f)
Big Plate 1993-4;
%FinalChanges3: End Here
 (g) PYTHON; (h) ARGO;
(i) MSAM (2-beam) - upper
 point uses entire data set, the
lower point has unidentified point sources removed;
 (j) MAX3 (GUM region); (k) MAX3 ($\mu$Pegasus
region) showing here  unidentified residual after dust
subtraction; and (l) MSAM (3-beam) - the upper point
uses the entire data set, lower point has unidentified point sources removed.
   }
\vspace{0.5cm}
%\vspace{-1.45cm}
\label{fig:f5}
 \end{figure}

\begin{figure}
%\centerline{\psfig{file=snowf4.ps,width=3.4in}}
\caption{
Effect of varying cosmological parameters on the power
spectrum over intermediate ($2<\ell \le 100$) scales.
The solid curve is the baseline spectrum,  $n_s=1$
(Harrison-Zel'dovich), $r=0$,
 h=0.5,
$\Omega_{\Lambda}=0$ and  $\Omega_B h^2 =.0125$.
The enveloping dashed  curves also have $n_s=1$, but other
cosmological parameters are varied within their
astrophysical bounds.  Dashed curves from top to
bottom: (1)~$\Omega_{\Lambda}$ increased to 0.6;
 (2) $h$ decreased to 0.4;
(3) $\Omega_B h^2$ induced to 0.025;
(4)~$\Omega_B h^2$ reduced to 0.005;
(5)~$h$ increased to 0.75.
Curves (2) and (3) are difficult to distinguish because
they overlap considerably.
 Note how this whole family of
curves is well-separated from the
lower dot-dashed curve which corresponds to $n=.85$ and $r=1$.
     } \label{fig:f4} \end{figure}

Fig.~4  and
Fig.~7 illustrate  important lessons concerning COBE DMR:
First, the initial emphasis on extracting the COBE quadrupole moment
and determining the spectral index has perhaps been  misplaced.
The quadrupole is difficult to extract empirically because of
the galactic background. It has limited use theoretically because
of
the large cosmic variance for this multipole (as shown in
Fig.~7).  Most important,
the quadrupole contains
no information that cannot be obtained by measuring the higher multipole
moments, which can be fixed empirically
and theoretically to greater
accuracy.
The second lesson is that measuring the spectral index using COBE DMR alone is
limited by the fact that
it is based on the detection
of the first 20-30 multipole moments only.
The range of $\ell$ is rather short for determining accurately the tilt
in the power spectrum
with increasing $\ell$.
  Fig.~4 suggests
that experiments at somewhat larger $\ell$ ($\gtrsim 50$), when combined with
COBE DMR, can produce a more accurate measure of the spectral
index because of the larger lever-arm in $\ell$.
There is the additional advantage that the larger $\ell$ multipole
moments have smaller cosmic variance.

\vspace*{.1in}
\noindent
{\bf Milestone 3: Observations of the First Doppler Peak}

  The existence of the first Doppler  peak
is a generic feature of inflationary models (and CDM or
HDM models of large-scale structure formation)  assuming standard
recombination or reionization at $z\lesssim 100$.
The position of the Doppler
peak, at $\ell \approx 220/\sqrt{\Omega_{total}}$, tests
whether the
 Universe is spatially flat, as inflation
predicts \cite{kamion}$^{-}$\cite{seljak}.
%\cite{kamion,hu,seljak}
The height of the first Doppler peak depends sensitively on
$h$, $\Omega_B$, and $\Omega_{\Lambda}$,   and on
the reionization history.   Discriminating the various
effects is difficult  since rather different choices of
cosmological parameters can lead to virtually indistinguishable
CMB anisotropy spectra.  We refer to this problem as {\it
cosmic confusion}.  Its implications are discussed in the
next section.

%40 lines
\begin{figure}
%\centerline{\psfig{file=gorski.ps,width=3.4in}}
\caption{
Cross-correlation between 53~GHz and 90~GHz frequency channels
in the COBE DMR 2-year map compared with three theoretical models:
(a)~flat CDM universe (solid curve); (b)~$\Lambda$-dominated
flat
universe with $\Omega_\Lambda =0.8$ (upper dashed curve);
(c) open  baryon-dominated universe with $\Omega=0.03$ (lower
dashed curve).  The cross-correlation has been expanded into
orthogonal functions on the cut sky labeled by the index $\ell$.
(Courtesy of K. Gorski.)
     } \label{fig:gorski} \end{figure}

Fig.~5 illustrates the
theoretical predictions and the
present experimental limits.
Error bars represent one-sigma bounds.
 It is difficult to draw any firm conclusions
from this range of observational results.  However, it is
interesting to note that a sequence of new MAX results re-examining
the GUM (Gamma Ursa Major) and two other regions  are
consistent with the large amplitude found by the earlier
MAX3 GUM experiment (see Fig.~8).  Present results suggest
a rather large Doppler peak.

Observation of the first Doppler peak is an especially important
discriminant between cosmic defect (strings, textures, etc.) and
inflationary/CDM models for large-scale structure formation.
Fig.~5 also illustrates the predictions for cosmic
texture models \cite{turok}. The predictions for other
cosmic defect models are similar.  The cosmic texture predictions shown
here  assume
reionization at $z\gtrsim 200$, which accounts partially for the
suppression of fluctuations just where inflation predicts a
Doppler peak. However, even if no reionization is assumed,
 cosmic defect models   normalized to COBE DMR predict smaller
amplitude fluctuations at $1^{\circ}$ scales than inflationary/CDM
models (which is related to why they require high bias parameters)
\cite{private}.
For both reasons, cosmic defect models
 generally predict a
signal considerably less than the Doppler peak predicted for
$n_s=1$ inflationary models, although precise calculations of
the anisotropy are not yet available for cosmic defect models
with standard recombination.

%30 lines HERETHEN
\begin{figure}
%\centerline{\psfig{file=snowf6.ps,width=4.0in}}
\vspace{0.0cm}
\caption{
Blow-up of the power spectrum over a  narrow range of $\ell$
near the  left (low $\ell$) slope of
the  first Doppler peak showing
recent MAX4 results  from the  $\iota$-Draconis,  GUM (Gamma Ursa
Major),  and $\sigma$-Hercules
regions.  The two theoretical curves are ($r=0$, $n_s=1$, solid)  and
($r=1$, $n_s=0.085$, dashed) inflationary/CDM predictions, respectively.
The dot-dashed curve shows the prediction for
cosmic textures (with reionization).
The data are:
(a) MAX4 GUM 6 cm$^{-1}$ (triangle) and 9 cm$^{-1}$ (diamond) channels;
(b) MAX4 $\iota$-Draconis 6 cm$^{-1}$ (triangle) and 9 cm$^{-1}$ (diamond)
channels;
(c) MAX4 $\sigma$-Hercules 6 cm$^{-1}$ (triangle) and 9 cm$^{-1}$ (diamond)
channels;
(d) MAX4 average (open circles) over all channels for (left-to-right) GUM,
 $\iota$-Draconis and  $\sigma$-Hercules;
(e) MAX4 GUM 3.5 cm$^{-1}$ for (left-to-right) GUM,
 $\iota$-Draconis and  $\sigma$-Hercules;
(f) MSAM2 (2-beam) - the upper point is based on the entire data set, the
  lower point
has unidentified point sources removed;
(g) MAX3 GUM; (h) MAX3  $\mu$ Pegasus (showing here  unidentified residual
after dust
subtraction).}
\vspace{0.5cm}
\label{fig:f6}
\end{figure}

%33 lines
\begin{figure}
%\centerline{\psfig{file=snowf7.ps,width=4.0in}}
\vspace{0.0cm}
\caption{
Power-spectrum beginning
from $\ell=50$ spanning all Doppler peaks and
the Silk damping region with experimental detections superimposed.
Error bars represent one-sigma detections and triangles represent
95\% confidence upper limits.
The two curves both correspond to pure scalar ($r=0$),
$n_s=1$ spectra with $h=0.5$.  The solid curve corresponds to the
standard value $\Omega_B h^2 =0.0125$ and the dot-dashed curve
corresponds to $\Omega_B h^2 = 0.025$.  Note that the second Doppler
peak is suppressed relative to the first and
second as $\Omega_B h^2$ increases.
 The
experiments correspond to:
%FinalChanges4: Begin Here
(a) South Pole 1991; (b) South Pole 1994; (c)
Big Plate 1993-4;
%FinalChanges4: End Here
(d) PYTHON; (e) ARGO;
(f) MSAM (2-beam) - the upper
 point uses the entire data set, the
lower point has unidentified point sources removed;
 (g) MAX3 (GUM region); (h) MAX3 ($\mu$Pegasus
region) showing here  unidentified residual after dust
subtraction; and (i) MSAM (3-beam) - the upper point
uses the entire data set, the
lower point has unidentified point sources removed;.
 (j) White Dish; and (k) OVRO7.
    }
\vspace{0.5cm}
 \label{fig:f7}  \end{figure}

\vspace{.1in}
\noindent
{\bf Milestone 4: Observations of Second and Higher Order Doppler Peaks}

If the second Doppler peak can be resolved to near cosmic
variance uncertainty, additional
constraints on $\Omega_B h^2$ can be extracted.  For the
standard value $\Omega_B h^2=0.0125$, there is a sizable second
peak, but this peak is increasingly suppressed as $\Omega_B h^2$
increases (see Fig.~9 and discussion in Section 6).
Also, the second and higher order Doppler peaks probe
sufficiently small scales  to  test whether the dark
matter is cold or a mixture of hot and cold.

An increasing challenge for observers as the range proceeds
to smaller angular scales is foreground
subtraction
 from sources and from the Sunyaev-Zel'dovich
effect.
Fig.~9 illustrates the few experimental results that presently
span this range.

\vspace{.1in}
\noindent
{\bf Milestone 5: Observations of Damping}  \\
\indent
If experiments successfully trace the predicted power spectrum through
the second Doppler peak,  there is already overwhelming
evidence in favor of inflation and dark matter models of
large-scale structure formation.   Observation of damping
at very small angular scales ($\lesssim 5'$)
due to photon diffusion (Silk damping \cite{silk}) and interference
through the thickness of the last-scattering surface is
corroboration of more subtle effects on the cosmic microwave
background \cite{Efst}.  If there was significant reionization (which
can already be  determined by experiments at larger
angular scales), then
experiments probing this
range might find evidence for
secondary perturbations.  Fig.~9 illustrates this range
and the present limit from OVRO7.

%FinalChanges6: Begin Here
\section{Cosmic Confusion??}

 The CMB anisotropy power spectrum entails thousands of
$C_\ell$'s and depends only upon a handful   of parameters,
($n_s$, $n_t$, $r$, $h$, $\Omega_B$, $\Omega_{\Lambda}$, $\ldots$).
One may have hoped that measurements of the power spectrum
would be able to test inflation and resolve independently
each of the parameters.  Unfortunately,
it is possible to continuously vary certain combinations of
the cosmological parameters without significantly changing
the power spectrum.
We refer to this  ``degeneracy'' as {\it cosmic
confusion}.\cite{yamada,Bond94}
  It means that CMB anisotropy experiments
can localize  a
hypersurface in cosmic parameter space
 but
the likelihood along that hypersurface
 will hardly vary.   This confounds efforts to separately
determine the values of cosmological parameters.

The degree of cosmic confusion depends on the range in $\ell$
over which the anisotropy spectrum is measured and the precision of
the measurement.  It seems possible that
measurements over the next decade will determine the
spectrum  from small $\ell$
through the first Doppler peak $\ell \sim 300$
with an error comparable to cosmic variance.
In this section, we will assume
that these observations can be made and
show that cosmic confusion is  a significant problem.
If precise observations are restricted to this limited range
of multipoles,
other cosmic observations must be combined with CMB anisotropy
measurements to break the degeneracy in fitting cosmological
parameters.  The degree of cosmic confusion can be reduced
if even more of the power spectrum can be determined precisely
through the second Doppler peak ($l\sim 500$ or
$sim$10 arcminute scales).  However, mapping the full sky at
such high resolution is an extraordinary challenge and, even if
one succeeds, it is possible that foregrounds will obstruct
measurements at these small angular scales.  Hence, it seems
reasonable, at least for the next decade,
to consider the more conservative case in which
precise measurements are made only through the first Doppler peak.

%FinalChanges6: End Here

It should be emphasized at the outset
 that extraordinarily valuable information can be
gained from microwave background measurements in spite of
cosmic confusion.  For example, as discussed under Milestone 2,
it is possible to test with high precision whether the
large-scale spectrum is scale-free and  measure the spectral
slope.  As shown in Fig.~6, there is negligible interference due to
 uncertainties in other
cosmological parameters.    Hence, it is certainly  possible to test
unambiguously some of the key predictions of inflation.

However,
one may have  hoped for more.  Some may argue that Harrison and
Zel'dovich  made the case for a scale-invariant spectrum
without invoking inflation \cite{Harr}.  In this case, verifying the
additional relations Eqs.~(\ref{rel1}-\ref{rel3}),  which have no natural
explanation from Harrison and Zel'dovich's point-of-view, would
be significant, added support for inflation (and
there is the obvious converse).  Also, it would have been
a tremendous boon if other cosmological parameters could be
unambiguously determined.  This is where cosmic confusion
disappoints.  All is not lost, though.
  In some cases (see, for example, the comments at the
end of this section), the degree of confusion is minor.
Even in the worst cases, powerful relations result when
CMB measurements are combined with other astrophysical measurements.

As an illustration of cosmic confusion, consider
 a baseline (solid line) spectrum $(r=0 | n_s=1,{\rm
h}=0.5,\Omega_{\Lambda}=0)$.  Increasing $\Omega_{\Lambda}$ (or
decreasing $h$) enhances small-angular scale anisotropy by
reducing the red shift $z_{\rm eq}$ at which radiation-matter equality
occurs. CMB anisotropy experiments can determine either $r | n_s$,
$\Omega_{\Lambda}$, or $h$  quite accurately if the other two
parameters are known \cite{Bond94}.
However, cosmic confusion arises if $r | n_s$,
$\Omega_{\Lambda}$ and $h$ vary
simultaneously.
 Fig.~10 shows spectra for models
lying in a two-dimensional surface of $(r|n_s,{\rm
h},\Omega_{\Lambda})$ which produce nearly identical spectra.  In one
case, $r | n_s$ is fixed, and increasing $\Omega_{\Lambda}$ is nearly
compensated by increasing $h$.  In the second case, $h$ is
fixed, but increasing $\Omega_{\Lambda}$ is nearly compensated by
decreasing $n_s$.
% [\kofn]

\begin{figure}
%\centerline{\psfig{file=snowf8a.ps,width=3.4in}}
\caption{
Examples of different cosmologies with nearly
identical power spectra and band powers.
The solid curve is $(r=0 | n_s=1, h=0.5, \Omega_{\Lambda}=0)$.
   The
other two curves explore degeneracies in the $(r=0 | n_s=1,{\rm
h},\Omega_{\Lambda})$ and $(r | n_s, h=0.5,\Omega_{\Lambda})$
planes.  In the dashed curve, increasing $\Omega_{\Lambda}$ is almost
exactly compensated by increasing $ h$.  For the dot-dashed curve,
the effect of changing to $r=0.42 | n_s=0.94$ is nearly compensated by
increasing $\Omega_{\Lambda}$ to 0.6.
%FinalChanges7: Begin Here
Restricting attention to the range from $\ell=2$ through the
first Doppler peak, one finds that
the curves are difficult to distinguish, especially given
cosmic variance (vertical hashing is one-sigma variance).
If precision measurements of the second Doppler peak or beyond
can be made, it should be possible to distinguish some of the
models.
%FinalChanges7: End Here
   } \label{fig:f8} \end{figure}

Further cosmic confusion arises if we consider ionization history.
Let $z_R$ be the red shift at which we suppose sudden, total
reionization of the intergalactic medium.  Fig.~11 compares spectra
with standard recombination (SR), no recombination (NR) and late
reionization (LR) at $z_R=50$, where $h=0.5$ and
$\Omega_{\Lambda}=0$.
Reionization or no recombination suppresses the small angular
scale anisotropy,
 which can be confused with a
decrease in $n_s$ (see figure).  Inflation-inspired models, {\it e.g.}, cold
dark matter models, are likely to have negligibly small $z_R$.

\begin{figure}
%\centerline{\psfig{file=snowf9.ps,width=3.4in}}
\caption{
Power spectra for models with standard recombination (SR), no
recombination (NR), and `late' reionization (LR) at $z=50$.  In all
models, $ h=0.5$ and $\Omega_{\Lambda}=0$.  NR or
reionization at $z \ge 150$ results in substantial suppression at
$\ell \ge 100$.  Models with reionization at $20 \le z \le 150$
give moderate suppression that can mimic decreasing $n_s$ or
increasing $h$; for example, compare the $n_s=0.95$ spectrum with SR
(black, dashed) to the $n_s=1$ spectrum with reionization at $z=50$
(grey solid).
}
\label{fig:f9}
\end{figure}

The results can be epitomized by some simple rules-of-thumb: Over the
$30'-2^{\circ}$ range, the $C_{\ell}$'s for fixed $\ell$
 are roughly
proportional to the maximum of $ C_\ell$ at the top of
 the first Doppler
peak.
 The maximum (corresponding to $\sim .5^{\circ}$ scales)
is exponentially sensitive
to $n_s$.  Since scalar fluctuations account for the Doppler peak, the
maximum decreases as the fraction of tensor fluctuations (or $r$) increases.
The maximum is also sensitive to
the red shift at matter-radiation equality (or, equivalently,
$(1-\Omega_{\Lambda})h^2$), and to the optical depth at last
scattering for late-reionization models, $\sim z_R^{3/2}$.  These
observations are the basis of an empirical formula (accurate to
$\lesssim 15$\%)
\begin{equation} \label{maxlaw}
\frac{C_\ell }{\avrg{ C_\ell }_{dmr}}
\Big\vert_{max}
\approx A \; e^{B \; \tilde{n}_s} \ ,
\end{equation}
with $A=0.15$, $B=3.56$, and
 \begin{equation}
\begin{array}{rcl}  \label{ntilde}
\tilde{n}_s &  \approx &  n_s -0.28 \, {\rm log}(1+ 0.8 r) \\ & &
-0.52 [(1-\Omega_{\Lambda}) h^2]^{\frac{1}{2}}\,
    - 0.00036 \, z_R^{3/2}+.26 \ ,
\end{array}
\end{equation}
where $r$ and $n_s$ are related by Eq.~(\ref{rel3})
for generic inflation models.
($\tilde{n}_s$ has been defined such that $\tilde{n}_s=n_s$ for $r=0,\;
h=0.5, \; \Omega_{\Lambda}=0,$ and $z_R=0$.)
Hence, the predicted anisotropy for any experiment
 in the range $10'$ and larger  is not
separately dependent on $n_s$, $r$, $\Omega_\Lambda$, etc.; rather, it
is function of the combination $\tilde{n}_s$.

Eq.~(\ref{maxlaw})
implies that the CMB anisotropy measurements are exponentially
sensitive to $\tilde{n}_s$.  Hence, we envisage that $\tilde{n}_s$
will be accurately determined in the foreseeable future.
Then, Eq.~(\ref{ntilde})
implies that
the values of the cosmological parameters are constrained to a
surface in parameter-space.
Cosmological models corresponding
to any point on this surface yield indistinguishable CMB anisotropy
power spectra
(thru the first Doppler peak).
To determine which point on the surface corresponds
to our Universe requires other astrophysical measurements.
For example, limits on the age of the Universe from globular clusters,
on
$h$ from Tully-Fisher techniques, on $n_s$ from galaxy and
cluster counts, and on $\Lambda$ from gravitational lenses all reduce
the range of viable parameter space to a considerable
degree. It is by this combination of
measurements that the CMB power spectrum can develop into a high
precision test of cosmological models \cite{Crit93,Knox}.

%FinalChanges8: Begin Here
The difficulty posed by cosmic confusion
depends on which way the experimental results break.
We have already stated that cosmic confusion can be lifted if
precise measurements can be made through the second Doppler peak
and beyond.  But it should  also be noted that, even if we are  limited to
the first Doppler peak only, there are situations in which
cosmic confusion is  significantly reduced.
%FinalChanges8: End Here
Consider our baseline spectrum,
$(r=0 | n_s=1,{\rm
h}=0.5,\Omega_{\Lambda}=0)$.  If measurements
indicate a first Doppler peak which is significantly
below the baseline value,
then there are numerous effects which might be responsible:
 tilt ($n_s<1$), increased gravitational wave contribution ($r$),
increased Hubble parameter ($h>0.5$),
decreased $\Omega_B<0.05$, reionization, or perhaps
cosmic defects.   On the other hand, if measurements indicate
a first Doppler peak that is significantly higher than the
baseline value, the causative effects might be:
upward tilt ($n_s>1$), cosmological constant $\Omega_{\Lambda}>0$,
decreased Hubble parameter ($h<0.5$), or increased $\Omega_B>0.05$.
Of these four possibilities, the first three are extremely unattractive
due to theoretical and observational constraints;
the data strongly  suggest $\Omega_B>0.05$ with all other parameters held
at their original values.
 The degree of real cosmic confusion is considerably less in the
second case.  Since several of the present experiments (including MAX)
suggest anisotropy somewhat greater than the baseline value, these
comments are especially timely.

\section{Polarization}

In the discussion thus far, we have focused on what can be learned from
the CMB anisotropy measurements based on the power spectrum only.
The power spectrum represents only the two-point temperature
correlation function.  From a CMB anisotropy map, one can  hope to
measure three- and higher-point correlation functions, for example,
to test for non-gaussianity of the primordial spectrum. Another
conceivable test is the CMB polarization.  Measurements of the
polarized temperature autocorrelation function \cite{Poln} and
of the cross-correlation between polarized and unpolarized
anisotropy \cite{turokp} are further tests of inflation.

\begin{figure}
%\centerline{\psfig{file=snowf10.ps,width=3.3in}}
\caption{
The percentage polarization in $\Delta T/T$  versus multipole
moment $\ell$ predicted for
  an inflationary model with $n_s=0.85$, $h=0.5$, cold
  dark matter, and standard
   recombination.  (For this value of
$n_s$, inflation predicts equal scalar and tensor
    contributions to the unpolarized quadrupole.)
The upper panel represents the prediction  for standard recombination and
the lower panel is for a model with no recombination.
}
\label{fig:f10}
\end{figure}

Calculations for realistic parameters, though,
suggest that the polarization is unlikely to be detected  or to
provide particularly useful tests of  cosmological parameters \cite{polar}.
% [\polar].
For example,  it had been hoped that large-scale (small $\ell$)
polarization measurements would be  useful for discriminating scalar and
tensor contributions to the CMB anisotropy,
%[\hope],
thereby
measuring $r$.  In the upper panel
of Fig.~12, we show the percentage polarization (in $\Delta T/T$) for
scalar and tensor modes for a model with $r=1$ and $n_s=.85$, an
example where there are equal tensor and scalar contributions to the
quadrupole moment.  The figure shows that, indeed, there is a dramatically
different polarization expected for scalar versus tensor modes for
small $\ell$.  However, the magnitude of the polarization is less than
0.1\%, probably too small to be detected in the foreseeable future.
On scales less than one degree ($\ell>100$), the total polarization rises and
approaches 10\%,  a more plausible target for detection. However, the
tensor contribution on these angular scales is negligible, so detection
does not permit us to distinguish tensor and scalar modes.  In fact,
the predictions are relatively insensitive to  the cosmological
model, a notable exception being the reionization history.
The lower panel of Fig.~12 illustrates the prediction for a model
with no recombination.  The overall level of polarization is increased.
The tensor contribution is suppressed relative to scalar, so polarization
remains a poor method of measuring $r$.   However, the polarization
at angular scales of a few degrees ($\ell \approx 50$) rises to nearly
5\%, perhaps sufficient for detection.  An observation of polarization at
these  angular scales would be consistent with a non-standard reionization
history.

Not only is the predicted polarization small in all cases, but little is
known about
what the foreground contamination will be.  At a minimum, the
foreground from dust, synchrotron, etc., interferes with measurements
at the 1\% level \cite{timbie}.   Polarization measurements  can provide
useful corroborating evidence or surprises.  For example, a
sizable polarization (20\%,  say) would be unexpected in
all models.  However, if inflation is correct, then (unpolarized) anisotropy,
rather than polarization,  appears to be the most useful
discriminant for the foreseeable future.

\section{Conclusions}

The next decade is sure to be a historic period in the
 endeavor to understand the origin and evolution of the
Universe.  In this article, we have focused only on
 measurements of the cosmic microwave background. We have shown
how these measurements alone will allow us to test the
inflationary hypothesis and place new constraints on
almost all cosmological parameters.
Table~2 below summarizes the specific sequence of five  milestones in
CMB anisotropy experiments that  need to be achieved to
accomplish these goals, the range of multipole moments ($C_{\ell}$)
that need to be probed, and which aspects of inflationary
predictions
is tested at each
milestone.
 Passing all milestones is overwhelming support for
inflation; failing to pass one or more milestones is either
invalidation or, at least,  indication of some significant,
additional surprise.

%FinalChanges9: Begin Here
The ability to  separately resolve inflationary and other cosmological
parameters from microwave background measurements alone is limited by
cosmic confusion, the phenomenon that different choices of cosmological
parameters result in virtually identical CMB
anisotropy spectra.
 Confusion is not a significant problem  if the observed anisotropy
at the first Doppler peak ($\sim 0.5^{\circ}$)
is found to be large or if precise measurements can be
extended to  the second Doppler peak and beyond
 (see remarks at the end of Section~8).
However, it is not clear whether either of these conditions will
be met.  Even in that case, it is still possible to use the CMB
anisotropy measurements as a powerful tool for determining cosmological
parameters, but only after combining CMB results with other
known astrophysical constraints.
%FinalChanges9: End Here

Another extraordinary series of efforts during this same decade
entails measurements of large-scale distribution
and velocities of galaxies.  These observations probe the
distribution of matter on scales generally smaller than but overlapping
the range probed by microwave background experiment.
The ultimate
goal is to find a simple, intuitive theory which joins together
the  microwave background and large-scale structure observations.

The present situation is summarized in Fig.~13, in which the
predictions of
 dark matter models (in linear approximation only)
are superimposed \cite{tk}.  (Non-linear corrections have been computed
for some models via numerical simulation.)
At present, the cold dark matter models are the simplest from
a theoretical point-of-view, but predict too much power
on small scales when normalized to COBE DMR.
Adding new dark matter components, as in mixed dark matter models, leads to
a better fit, but at the cost of introducing the special condition
of nearly equal hot and cold matter densities that is difficult
to explain from microphysics.  Tilting the spectrum  to  reduce
power at small scales is naturally incorporated into inflation,
but the
numerical simulations do not fit large-scale structure as successfully.
An interesting puzzle appears to be brewing.

Microwave background and large-scale structure observations are
dramatic examples of the transformation of cosmology from
metaphysics to hard science.
These, combined with the more classical cosmological measurements
of $h$, $\Omega$, and $\Lambda$,
 are rapidly evolving into  highly precise tools for testing
theories and measuring fundamental parameters.
As the new millennium commences, cosmological science stands
at a critical  crossroads: we do not know whether
 the Universe can be explained solely on the basis of physical
laws that can be tested locally, or whether  knowledge of initial
conditions or new physics beyond our grasp is required.
Evidence that the Universe is scale-free ({\it e.g.}, spatially flat,
scale-free primordial spectrum of fluctuations) is consistent
with the notion that the physical laws  governing the Universe
involve microphysical scales only,
as suggested by  present
understanding.  Evidence that the Universe is measurably open
(or closed) or that there are large-scale features in the
primordial spectrum suggests initial conditions or coincidences
or new physical laws that may not be probed by other means.
The series of measurements anticipated in the next decade may
determine the future path of   cosmology,
and, thereby, the ultimate limitations on human comprehension of
the Universe.

%44 lines to move
\begin{table}[t]
\begin{center}
\begin{tabular}{||l|c|c||}
\hline \hline
& &  \\
 Milestone & Range of $\ell$   & What It Tests  \\
 & & \\
\hline \hline & & \\
1. Large Scale Fluctuations & $2 \lesssim \ell \lesssim 30$ &
Spectral Amplitude
\\ & & \\
\hline
 & & \\
2.  Plateau at Intermediate Scales  & $10 \lesssim \ell \lesssim 100$
& Spectral shape/slope
\\ & &
 \\
\hline
& & \\
3.  First Doppler Peak & $100 \lesssim \ell \lesssim 300$ &  \\
& & \\
\hspace*{3em}  a. Value of $\ell$ at the maximum &  & Flatness   \\
& & \\
\hspace*{3em}  b. Height &  & Constraints on $h$, $\Omega_B$,
\\
 & & $\Omega_{\Lambda}$
 and reionization? \\
\hline  & & \\
4. Second \&  & $300 \lesssim \ell \lesssim 800$ &
 Constraints on
	$\Omega_B h^2$,  \\
Higher Doppler Peaks & & CDM vs. MDM \\ & & \\
\hline  & & \\
5. Damping & $\ell \gtrsim 1000$ & Silk Effect, \\ & &  Cosmo. parameters?
 \\ & & \\ \hline \hline
\end{tabular}
\vspace{0.5cm}\end{center}
\caption{
Five Milestone Tests of Inflation and Dark Matter Models of
Large-scale Structure }
\label{tab:mile}\vspace{0.5cm}
\end{table}

\begin{figure}
%\centerline{\psfig{file=snowf11.ps,width=3.3in}}
\caption{
Comparison of matter density power spectra for cold dark matter (CDM)
tilted cold dark matter (TCDM), hot dark matter (HDM) and mixed
cold and hot dark matter models (MDM) of largescale structure
formation.  All  theoretical curves are normalized to COBE DMR and
include only linear approximation.  (Non-linear corrections become
important at small angular scales ($\le 10$~Mpc).
    }
\label{fig:f11}
\end{figure}

\setcounter{secnumdepth}{0} %this ensures that there are no section numbers
                            %from here on in the text. Don't remove.

\section{Acknowledgements}

Many  results in this paper  are drawn from
collaborations with
R. Crittenden,  J. R. Bond, R. Davis, G. Efstathiou, H. Hodges,
G. Smoot, and M. S. Turner.
I thank  P. Lubin,  P. Meinhold  and J.
Gundersen for showing the South Pole 1994 results prior to
publication and  discussions on MAX data,
and K. Gorski for providing  Fig.~7.
I thank the Snowmass Workshop organizers,
Roberto Peccei and  Rocky Kolb, and the participants in the
Cosmology working group for their interest,  encouragement and
support.
  I also thank Cynthia Sazama and her staff for their
outstanding assistance (and the miraculous weather at the final,
mountaintop banquet) .
The author is a John Simon Guggenheim Fellow  during 1994-5.
This research was supported by the DOE at Penn (DOE-EY-76-C-02-3071);
 and by National Science Foundation Grant
NSF PHY 92-45317 and  Dyson Visiting Professor Funds at the Institute
for Advanced Study.

\vspace*{.1in}
\appendix{{\bf Appendix: Flat Band Power Estimates}}

The observational results reported
in Figs.~4, 5, 8 and 9
 represent
``flat band power estimates" derived for each experiment.
In this appendix, we explain the  reasoning behind
band power estimation; we briefly discuss the
particular cases of GACF (gaussian auto-correlation function) and
flat band power estimation, and approximate methods for converting
from one estimate to another.

 Each experiment, depending on the geometry, scanning
strategy and
detectors, has different sensitivity to the $C_{\ell}$'s  which
can be expressed through a filter function \cite{wind}, $W_{\ell}$.
A number which can be directly extracted from experiment is the
rms variation in temperature,
\begin{equation} \label{rms}
(\Delta T/T)_{rms}^2 \equiv
\sum_{\ell} \; \frac{\ell+\frac{1}{2}}{\ell(\ell+1)}
\left[\frac{1}{2 \pi} \ell(\ell+1) C_{\ell}^{sky}\right] W_{\ell};
\end{equation}
the square brackets enclose the expression for
 the power spectrum of the  real
sky described multipole moments $C_{\ell}^{sky}$.
It is difficult to compare $(\Delta T/T)_{rms}$ directly with theoretical
predictions for three reasons: (1)~the value of $(\Delta T/T)_{rms}$ is
dependent upon the normalization and shape of the filter function, $W_{\ell}$;
(2)~a significant fraction of $(\Delta T/T)_{rms}$  is noise; and, (3)~the
value of $(\Delta T/T)_{rms}$ itself gives no information about the functional
form of the real sky power spectrum (that is, how $\ell(\ell+1) C_{\ell}^{sky}$
varies with $\ell$).

A first improvement is to compute the  rms band power, defined by
\begin{equation}\label{bp}
\frac{1}{2 \pi} \bar{\ell} (\bar{\ell} +1) C_{\bar{\ell}} \equiv
\frac{(\Delta T/T)_{rms}^2}{\sum_{\ell}  \frac{\ell +\frac{1}{2}}
{\ell (\ell+1)}
 W_{\ell}},
\end{equation}
where
\begin{equation}
\bar{\ell} \equiv \frac{\sum_{\ell}
\frac{\ell+\frac{1}{2}}{\ell+1}
W_{\ell}}{\sum_{\ell}\frac{\ell+\frac{1}{2}}{\ell
(\ell+1)}W_{\ell}}
\end{equation}
is the mean value of $\ell$ over the filter function band.
(Some authors also  delineate span of $\ell$ corresponding to the
half-width of the filter function by
a horizontal error bar;\cite{Bondonly,capri} we have not do so in the interests
of clarity.)
Band power is independent of the normalization of $W_{\ell}$, and only
weakly dependent on its shape.  However, problems~(2) and~(3) remain.

A second step is to introduce  ``$\langle{\it functional form}\rangle$ band
power estimation,''
a fit to the data assuming some particular functional form for
the power spectrum.  The assumed functional form could be the prediction
of some specific theoretical model, such as  the spectra shown in Fig.~2;
however, there are too many weakly constrained parameters in these models
at present.  Hence, a less biased approach for now is
to  assume a simpler form, such as gaussian (GACF)
or flat, which depends on only  a few free parameters. A maximum likelihood fit
of the data to the functional form
fixes the best-fit choices of the   parameters.
The  $\langle{\it functional form}\rangle$
 band power estimate is then:
\begin{equation}\label{ffbp}
\frac{1}{2 \pi} \bar{\ell} (\bar{\ell} +1) C_{\bar{\ell}} \equiv
\sum_{\ell} \; \frac{\ell+\frac{1}{2}}{\ell(\ell+1)}
\left[\frac{1}{2 \pi} \ell(\ell+1) C_{\ell}^{bf}\right] W_{\ell},
\end{equation}
where $\ell(\ell+1) C_{\ell}^{bf}/2 \pi$  has the assumed functional
form with  free parameters fixed at the best-fit values.
 Best-fitting   the data to a functional form
 which  resembles the real sky has the advantage that it filters out
noise.  By including parameters which change the shape of the assumed
power spectrum, some shape information can be extracted.

 A ``flat band power estimate" assumes
$\ell(\ell+1) C_{\ell}^{bf}/2 \pi$ is $\ell$-independent; the one free
parameter is the amplitude.
A gaussian autocorrelation function (GACF) estimate assumes that the
correlation function is gaussian:
$$
\begin{array}{rcl}
C(\theta)& =& \left\langle{\frac{\Delta T}{T}({\bf x}) \frac{\Delta T}{T}({\bf
x
'}) }
\right\rangle \\ &  =  &
  C(0) \,  \frac{\theta_c^2}{\theta_c^2+ 2 \sigma^2}
\, {\rm exp} \{- \theta^2/[2(\theta_c^2+ 2 \sigma^2)\}
\end{array}$$
where $\sigma$ is the FWHM beamwidth of the experiment divided by
$\sqrt{8 \;{\rm ln}\,2}$.
This corresponds to a gaussian power spectrum
  $\ell(\ell+1) C_{\ell}^{bf}/2 \pi$.
The spectrum has two free parameters,
the correlation
angle $\theta_c$ and the best-fit  amplitude $C(0)$ for the
given $\theta_c$.  (Some experiments report two-parameter best-fits.  Others
report single-parameter fits to $C(0)$ for fixed $\theta_c$.)

Ultimately, a likelihood fit over a spectrum of theoretical models is
the proper way to determine the best-fit model.
However, it is difficult
to show widely disparate models and many different
parameters  within a single likelihood plot.  It would be
highly desirable if one could somehow find unbiased  band power estimates
extracted from the experiment that can be compared
directly to the theoretical power spectra.

After testing various forms,
Bond {\it et al.} \cite{Bondonly,capri} have demonstrated that
flat band power estimates are a good choice for direct comparison with
theory, and this is what has been applied in  Figs.~4, 5,~8 and~9.
If the real sky  power spectrum is similar to one of the theoretical
curves, the flat band power approximation is  clearly a better
than  a gaussian (GACF)  one for the
plateau region (shown in Fig.~4); flat band power estimatation
 works surprisingly well for
the large $\ell$'s as well.  For example, if one computes
flat band power estimates  assuming the theoretical predictions,
the estimates closely hug the theoretical curves even through
the Doppler peaks \cite{capri}.

As described above,
the proper method of obtaining the flat band power estimate is
through a full, likelihood fit to the data assuming the flat spectrum.
At this point, though, many experimental groups only publish a
GACF analysis.
The following five easy
steps are an approximate conversion formula works well for
most experiments.  From a reported $\theta_c$ and $\Delta T/T_{GACF}$:
\begin{enumerate}
\item Compute $\ell_c \equiv 1/[2 {\rm sin}(\theta_c/2)]$.
\item Compute the filter function $W_{\ell}$  for the given experiment.
\item Compute $$N\equiv \sum_{\ell}\frac{\ell+\frac{1}{2}}{\ell(\ell+1)}
\left(\frac{\ell+\frac{1}{2}}{\ell_c+\frac{1}{2}}\right)^2 {\rm exp}\left[
-\frac{1}{2} \left(\frac{\ell+\frac{1}{2}}{\ell_c+\frac{1}{2}}\right)^2\right]
W_\ell$$
\item Compute $$D= \sum_{\ell}\frac{\ell+\frac{1}{2}}{\ell(\ell+1)}W_\ell$$
\item Then the flat band power estimate for $\ell(\ell+1)C_{\ell}$ is
$$\frac{\bar{\ell} (\bar{\ell}+1) C_{\bar{\ell}}}{2 \pi}=
\left(\frac{\Delta T}{T}\right)_{GACF}^2 \frac{N}{D}$$.
\end{enumerate}

%FinalChanges5: Begin Here

%FinalChanges5: End Here

\end{document}